\documentclass[aps,twocolumn,preprintnumbers,amsmath,amssymb,superscriptaddress,longbibliography]{revtex4-1} 

\usepackage{graphicx}
\usepackage{dcolumn}
\usepackage{color}
\usepackage{amsmath}
\usepackage[breaklinks,colorlinks,bookmarks=false,citecolor=blue,linkcolor=red,urlcolor=blue]{hyperref}

\newcommand{\transfer}[2]{ | {#1}\rangle\langle {#2} |}

\begin{document}

\preprint{\href{https://doi.org/10.1103/PhysRevB.98.174432}{Phys. Rev. B {\bfseries 98}, 174432 (2018)}}

\title{Thermodynamic properties of the Shastry-Sutherland model\\
from quantum Monte Carlo simulations}

\author{Stefan Wessel}
\affiliation{Institut f\"ur Theoretische Festk\"orperphysik, JARA-FIT and 
JARA-HPC, RWTH Aachen University, 52056 Aachen, Germany}

\author{Ido Niesen}
\affiliation{Institute for Theoretical Physics and Delta Institute for 
Theoretical Physics, University of Amsterdam, Science Park 904, 1098 XH 
Amsterdam, The Netherlands}

\author{Jonas Stapmanns}
\affiliation{Institut f\"ur Theoretische Festk\"orperphysik, JARA-FIT and 
JARA-HPC, RWTH Aachen University, 52056 Aachen, Germany}

\author{B.~Normand}
\affiliation{Neutrons and Muons Research Division, Paul Scherrer Institute, 
5232 Villigen-PSI, Switzerland}

\author{Fr\'ed\'eric Mila}
\affiliation{Institute of Theoretical Physics, Ecole Polytechnique F\'ed\'erale
Lausanne (EPFL), 1015 Lausanne, Switzerland}

\author{Philippe Corboz}
\affiliation{Institute for Theoretical Physics and Delta Institute for 
Theoretical Physics, University of Amsterdam, Science Park 904, 1098 XH 
Amsterdam, The Netherlands}

\author{Andreas Honecker}
\affiliation{Laboratoire de Physique Th\'eorique et Mod\'elisation, CNRS 
UMR 8089, Universit\'e de Cergy-Pontoise, 95302 Cergy-Pontoise Cedex, France}

\date{August 6, 2018; revised October 30, 2018}

\begin{abstract}
We investigate the minus-sign problem that afflicts quantum Monte Carlo (QMC) 
simulations of frustrated quantum spin systems, focusing on spin $S = 1/2$, 
two spatial dimensions, and the extended Shastry-Sutherland model. We show 
that formulating the Hamiltonian in the diagonal dimer basis leads to a sign 
problem that becomes negligible at low temperatures for small and intermediate 
values of the ratio of the inter- and intra-dimer couplings. This is a 
consequence of the fact that the product state of dimer singlets is the 
exact ground state both of the extended Shastry-Sutherland model and of a 
corresponding ``sign-problem-free'' model, obtained by changing the signs of 
all positive off-diagonal matrix elements in the dimer basis. By exploiting 
this insight, we map the sign problem throughout the extended parameter space 
from the Shastry-Sutherland to the fully frustrated bilayer model and compare 
it with the phase diagram computed by tensor-network methods. We use QMC to 
compute with high accuracy the temperature dependence of the magnetic specific 
heat and susceptibility of the Shastry-Sutherland model for large systems up 
to a coupling ratio of 0.526(1) and down to zero temperature. For larger 
coupling ratios, our QMC results assist us in benchmarking the evolution 
of the thermodynamic properties by systematic comparison with exact 
diagonalization calculations and interpolated high-temperature series 
expansions.

\end{abstract}

\maketitle

\section{Introduction}

Frustrated quantum magnets, meaning those in which local exchange processes 
are in competition, are known to host a rich variety of physical phenomena 
within unconventional ground states ranging from various kinds of 
valence-bond crystal to quantum spin liquids \cite{Richter2004,Balents10,
HFMbook,DIEPbook}. However, the investigation of frustrated quantum spin 
models constitutes a real challenge, because there exist in general no unbiased 
methods by which to study their properties on sufficiently large lattices and 
at appropriately low temperatures. In two dimensions, quantum Monte Carlo 
(QMC) is the method of choice for studying the thermal properties of 
unfrustrated systems such as the square-lattice quantum antiferromagnet 
\cite{PhysRevLett.80.2705,HTN98}. In frustrated models, QMC suffers from a 
very severe ``minus-sign'' problem when performed in the standard basis of 
spin configurations, making it essentially impossible to obtain accurate 
results for any temperatures significantly below the typical coupling 
constants, which unfortunately constitute the only regime of interest 
in the context of exotic quantum physics. 

Two paradigmatic two-dimensional (2D) spin-1/2 frustrated models with 
approximate experimental realizations are the kagome antiferromagnet and 
the Shastry-Sutherland model \cite{ShaSu81}, the first as a candidate 
quantum spin liquid \cite{Liao17} and the second because of the 
remarkable, and still hotly debated, series of magnetization plateaus 
observed in SrCu$_2$(BO$_3$)$_2$ 
\cite{PhysRevLett.82.3168,Onizuka00,Kodama02,Sebastian08,Takigawa2011, 
Takigawa12,Jaime12,PhysRevLett.111.137204,ncomms16}. Both models have 
triangles as their building blocks, and hence a severe QMC sign problem. 
There is, however, also an important difference between them. While the 
ground state of the spin-1/2 kagome antiferromagnet is still highly 
controversial, the ground state of the Shastry-Sutherland model has been 
known for nearly 40 years \cite{ShaSu81,AM96,MiUeda99}. This model was 
actually constructed by Shastry and Sutherland as a 2D generalization of 
the spin-1/2 Majumdar-Ghosh chain \cite{MG69a,MG69b,Majumdar70}, 
i.e.~explicitly to have a product state of dimer singlets as the ground 
state. It seems logical to expect that knowledge of the ground state 
should help very significantly in investigating the low-temperature 
thermodynamics, but to date this has not been the case. Interpretation of 
the temperature dependence of the magnetic susceptibility 
\cite{PhysRevLett.82.3168} and specific heat of SrCu$_2$(BO$_3$)$_2$ 
\cite{Kageyama2000}, the nearly exact realization of the 
Shastry-Sutherland model, still relies primarily on exact diagonalization 
(ED) results \cite{MiUeda99,MiUeda00,MiUeda03} obtained for small 
lattices of up to only 20 sites \footnote{The magnetic susceptibility, 
$\chi(T)$, has also been analyzed by series expansions 
\cite{MiUeda99,PhysRevB.60.6608, PhysRevLett.85.3958}, but these are 
accurate only for temperatures above the maximum of $\chi$.}.

In this paper, we show that knowing the exact ground state of the 
Shastry-Sutherland model is indeed a considerable advantage, provided 
that one formulates QMC simulations in the dimer basis 
\cite{rus1,Alet16,NgYang17, rus3,rus4} rather than the conventional site 
basis. Unlike a number of fully frustrated models studied recently, in 
which the minus sign disappears completely in the dimer basis, we 
illustrate the extent to which the sign problem is still present 
throughout the generalized phase diagram that connects the 
Shastry-Sutherland model to the fully frustrated bilayer. Our key result 
is that, as long as the product of dimers is not only the ground state of 
the model itself, but also of the ``sign-problem-free'' model obtained by 
changing the sign of the positive off-diagonal matrix elements in the 
dimer basis, the sign problem decreases at low temperatures and 
disappears completely at zero temperature. From this insight we 
demonstrate using the example of the Shastry-Sutherland model that it is 
in fact possible to perform efficient QMC simulations to study the 
thermodynamics of certain frustrated quantum systems.

Our manuscript is organized as follows. In Sec.~\ref{The models} we 
introduce the Shastry-Sutherland model, the sign-problem-free counterpart 
model that provides insight into the nature of the minus-sign problem, 
and the extended model that interpolates between the Shastry-Sutherland 
case and the fully frustrated bilayer model, which enables us to discuss 
the ground-state phase diagram. In Sec.~\ref{The minus sign}, we exploit 
the sign-problem-free model to investigate the minus sign in the 
Shastry-Sutherland model by simulations in the dimer basis, from which we 
show how the sign problem is suppressed at low temperature in a large 
portion of the singlet-product phase. In Sec.~\ref{Physical results}, we 
build on this observation to compute the low-temperature specific heat 
and susceptibility of the Shastry-Sutherland model with high accuracy up 
to the critical coupling ratio. Section \ref{Conclusions} contains a 
brief summary and perspective. Two appendices provide details of 
tensor-network and high-temperature series-expansion methods, which we 
use to augment and benchmark our QMC analysis.

\section{The models}
\label{The models}

\begin{figure}[t!]
\hbox{\hbox{\includegraphics[width=0.49\columnwidth]{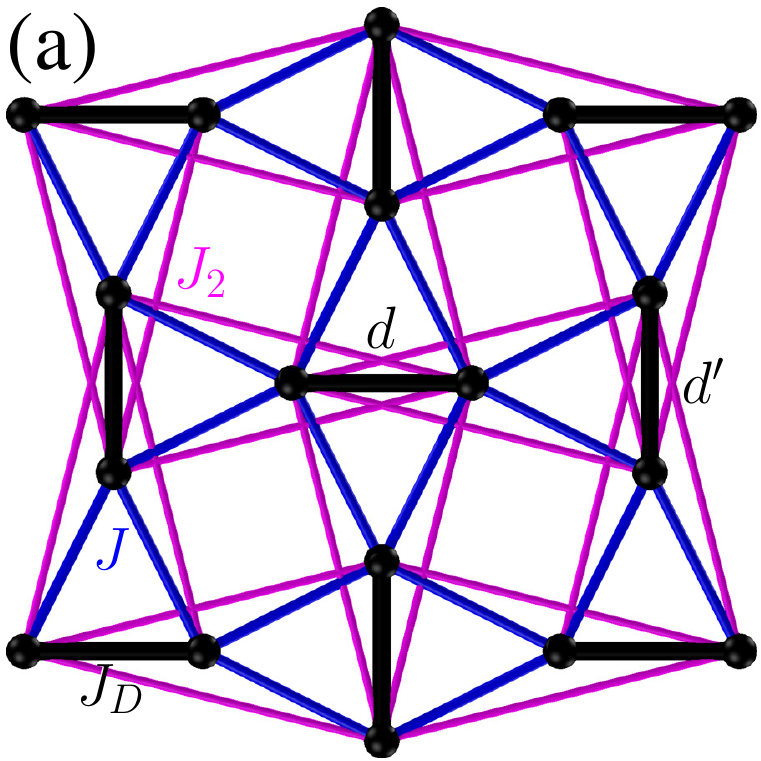}}\hfil%
\raise 10 pt\hbox{\includegraphics[width=0.49\columnwidth]{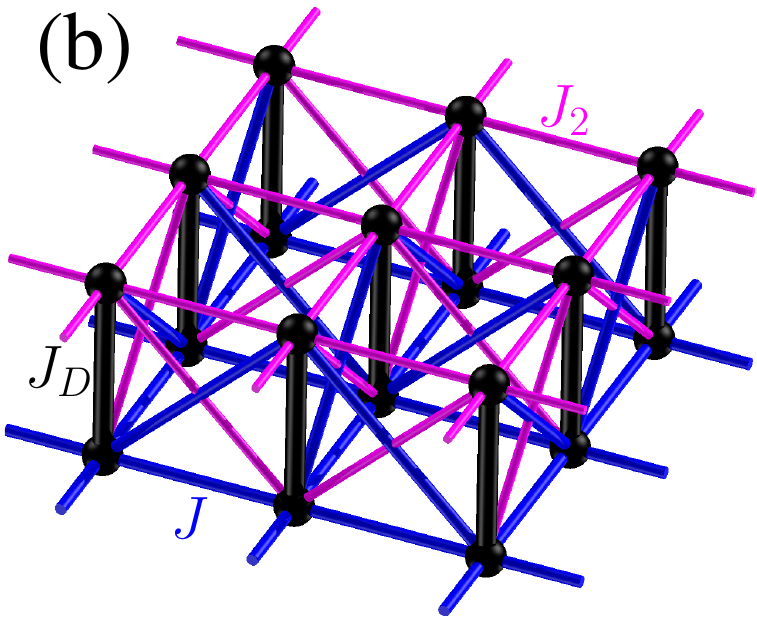}}}
\caption{Schematic representations of the extended Shastry-Sutherland model
[Eq.~(\ref{extended})] in (a) single-plane and (b) bilayer format.}
\label{fig:model}
\end{figure}

The Shastry-Sutherland model \cite{ShaSu81}, also known as the orthogonal 
dimer model \cite{MiUeda03}, is defined by the Hamiltonian
\begin{equation}
H = J_D \sum_{\langle i,j \rangle} \vec S_i \cdot \vec S_j + J \!\! \sum_{\langle 
\langle i,j \rangle\rangle} \!\! \vec S_i \cdot \vec S_j,
\label{ess}
\end{equation}
where $J_D$ is the intra-dimer coupling (denoted by $\langle ij \rangle$) 
and the inter-dimer coupling ($\langle\langle ij \rangle\rangle$), $J$, 
defines a square lattice as shown in Fig.~\ref{fig:model}(a). For small and 
intermediate coupling ratios, $J/J_D$, the ground state is an exact product 
of singlets formed on the dimer bonds~\cite{ShaSu81}. 

This is a property that the Shastry-Sutherland model shares with the fully 
frustrated $S = 1/2$ bilayer square lattice \cite{LinShen00,LinShik02,RDK06,
DRHS07,DKR10,rus4}. Because the sign problem is completely absent in the fully 
frustrated bilayer \cite{Alet16,NgYang17,rus4}, we consider an extended model
\cite{PhysRevB.60.6608,PhysRevLett.84.1808} defined by the Hamiltonian
\begin{equation}
H_\text{ext} = H + J_2 \!\!\! \sum_{\langle \langle \langle i,j\rangle\rangle\rangle} 
\!\!\! \vec S_i \cdot \vec S_j,
\label{extended}
\end{equation}
in which the addition of the next-neighbor inter-dimer coupling, $J_2$, 
illustrated in 
Fig.~\ref{fig:model}, interpolates between the Shastry-Sutherland model 
at $J_2 = 0$ and the fully frustrated bilayer when $J_2$ and $J$ are equal. 

\begin{figure}[t!]
\includegraphics[width=\columnwidth]{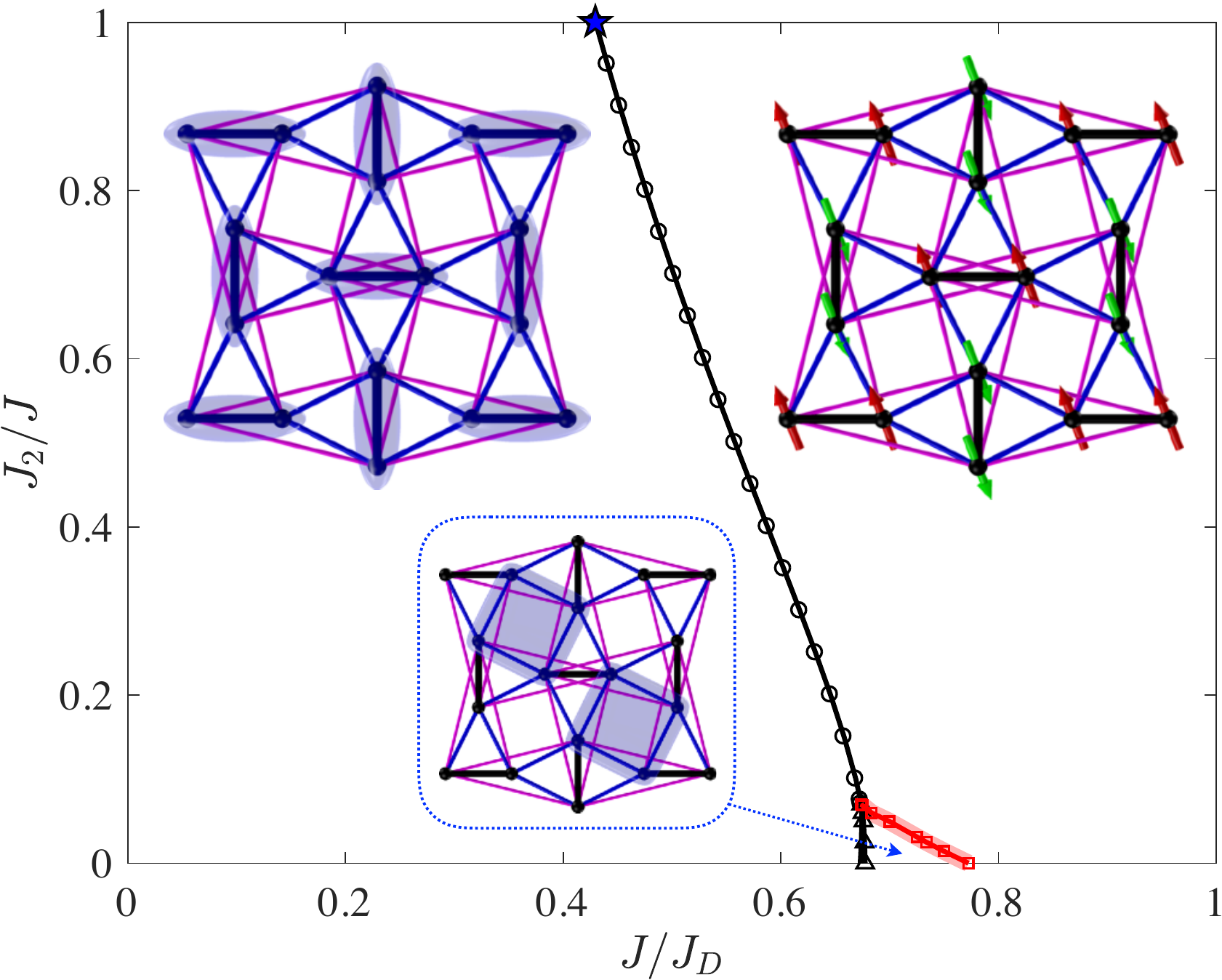}
\caption{Ground-state phase diagram of the extended Shastry-Sutherland 
model of Eq.~(\ref{extended}), obtained from iPEPS calculations. All phase 
transitions are first-order. The star at $(J/J_D, J_2/J) = (2.3279(1),1)$ 
denotes the location of the quantum phase transition in the fully frustrated 
bilayer, taken from Ref.~\cite{PhysRevLett.84.1808}. Insets show schematic 
representations of the dimer singlet-product phase (upper left), the 
square-lattice antiferromagnetic phase (upper right), and the intermediate 
plaquette phase (lower).} 
\label{fig:phase_diagram_peps}
\end{figure}

As we will show in Secs.~\ref{The minus sign} and \ref{Physical results}, 
the thermodynamic properties of both models can be studied very 
accurately by QMC as long as the interdimer couplings, $J$ and $J_2$, are 
not too large compared to $J_D$. As an aid to interpreting these results, 
we first obtain the full zero-temperature phase diagram of the extended 
model, $H_\text{ext}$, for which it is sufficient to consider $0 \le J_2 
\le J$. We apply the variational tensor-network approach of infinite 
projected entangled pair states (iPEPS), the technical details of which 
we provide in App.~\ref{sec:appipeps}. This method has been shown 
previously \cite{CorbozMila13} to provide very accurate results for the 
Shastry-Sutherland model [Eq.~(\ref{ess})], and in 
Fig.~\ref{fig:phase_diagram_peps} we show the phase diagram of the 
extended model [Eq.~(\ref{extended})]. The ground state is clearly a 
dimer-singlet phase at small inter-dimer couplings and a square-lattice 
antiferromagnetic phase at large $J$; this latter phase becomes an 
effective $S = 1$ square-lattice antiferromagnet in the bilayer limit 
($J_2/J = 1$) \cite{rus4}. Only near the opposite (Shastry-Sutherland) 
limit does a small regime of a third phase appear, the intermediate 
``plaquette'' phase (inset, Fig.~\ref{fig:phase_diagram_peps}) based on 
alternating squares of the $J$ lattice \cite{rkk,rtkk,LWS02,CorbozMila13}. 
The dimer and plaquette phases are gapped and all phase transitions are 
first-order. We comment that a previous investigation 
\cite{PhysRevLett.84.1808} came to very similar conclusions, 
except that it missed the intermediate plaquette phase.

With a view to our QMC calculations, we next define the sign-problem-free 
Hamiltonian corresponding to the extended spin model of Eq.~(\ref{extended}).
Working in the dimer basis, we change the signs of the off-diagonal matrix 
elements in such a way that all of them are non-positive. For a given dimer 
($J_D$) bond, $d$, we combine the two spins that form this dimer, $\vec S_{d,1}$ 
and $\vec S_{d,2}$, to introduce the total-spin operator, $\vec T_d = \vec 
S_{d,1} + \vec S_{d,2}$, and the spin-difference operator, $\vec D_d = \vec 
S_{d,1} - \vec S_{d,2}$. In defining $\vec D_d$ it is necessary to fix a 
convention regarding the assignment of the spin labels $1$ and $2$, and here 
we allocate $\vec S_{d,1}$ to the left (lower) spin on a horizontal (vertical) 
dimer in Fig.~\ref{fig:model}(a). By considering the local spin-singlet and 
-triplet states on dimer $d$, 
\begin{eqnarray}
|S \rangle_{d} & = & {\textstyle \frac{1}{\sqrt{2}}} ( |\!\uparrow \downarrow
\rangle_{d} - |\! \downarrow \uparrow \rangle_{d} ), \nonumber \\
|0 \rangle_{d} & = & {\textstyle \frac{1}{\sqrt{2}}} ( |\!\uparrow \downarrow 
\rangle_{d} + |\! \downarrow \uparrow \rangle_{d} ), \nonumber \\
|+ \rangle_{d} & = & |\! \uparrow \uparrow \rangle_{d}, \; 
|- \rangle_{d} = |\! \downarrow \downarrow \rangle_{d},
\label{eq:sdb}
\end{eqnarray}
we summarize the action of the total-spin and spin-difference operators in 
Table \ref{tab:matele}, where we use the conventional definitions $T_d^{\pm}
 = T_d^x \pm i T^y_d$ and $D_d^{\pm} = D_d^x \pm i D^y_d$.

\begin{table}[b]
\begin{tabular}{ c || c | c | c | c | c | c | c }
& $\vec{T}^2_{d}$ & $T^z_{d}$ & $T^+_{d}$       &  $T^-_{d}$        &  $D^z_{d}$  
& $D^+_{d}$        & $D^-_{d}$ \\
\hline
$| S\rangle_d$ & 0     & 0  & 0   &  0  &  $| 0 \rangle_{d}$   & $-\sqrt{2}|
 + \rangle_{d}$            & $\sqrt{2}| - \rangle_{d}$ \\
$| 0 \rangle_{d}$ & 2           & 0     & $\sqrt{2}| + \rangle_{d}$          
&  $\sqrt{2}| - \rangle_{d}$    &  $| S \rangle_{d}$   & 0       & 0 \\
$| + \rangle_{d}$ & 2     & 1     & 0           &  $\sqrt{2}| 0 \rangle_{d}$ 
&  0        & 0            & $-\sqrt{2}| S \rangle_{d}$ \\
$| - \rangle_{d}$ & 2           & $-1$    & $\sqrt{2}| 0 \rangle_{d}$    &  0  
&  0          & $\sqrt{2}| S \rangle_{d}$            & 0 \\
\end{tabular}
\caption{Action of total-spin and spin-difference operator components on the 
local dimer-basis spin states. Because $\vec{T}^2_{d}$ and $T^z_{d}$ are diagonal 
in this basis, we give only the eigenvalues for these operators. Note that in 
this basis $D^z_d$ is not diagonal.}
\label{tab:matele}
\end{table}

The Hamiltonian $H_{\mathrm{ext}}$ (\ref{extended}) consists of (i) a sum of the 
separate local couplings, i.e.~$J_D$, within each dimer $d$, which one may 
denote $H_d$, and (ii) sums over the inter-dimer terms, with couplings $J$ 
and $J_2$, that connect two neighboring orthogonal dimers. The local 
contribution may be expressed as $H_d = {\textstyle \frac12} \, \vec{T}_d^2
 - {\textstyle \frac{3}{4}}$, i.e.~in terms only of total-spin operators. 
The inter-dimer coupling for the two dimers $d$ and $d'$ indicated in 
Fig.~\ref{fig:model}(a) takes the form
\begin{equation}
H_{dd'} = {\textstyle \frac12} (J + J_2) \vec T_d \cdot \vec T_{d'} - {\textstyle 
\frac12}(J - J_2) \vec T_d \cdot \vec D_{d'}. 
\end{equation}
Clearly in the special case $J_2 = J$ the $TD$-coupling terms vanish, and in 
this limit, which corresponds to the fully frustrated bilayer model, QMC 
simulations formulated in the spin-dimer basis can be performed with no 
sign problem \cite{rus1,Alet16,NgYang17,rus3,rus4} despite the extreme 
frustration. By contrast, whenever $J_2 \neq J$, a finite $TD$-term is present 
in addition to the $TT$-terms, and in particular for the Shastry-Sutherland 
model ($J_2 = 0$) it is strong. This term leads to the reappearance of a 
minus-sign problem in the dimer basis, and in Sec.~\ref{The minus sign} 
we examine its severity in detail.

To complete the construction of the sign-problem-free Hamiltonian, 
$\widetilde{H}$, we start from $H_\mathrm{ext}$ and change the signs of all 
positive off-diagonal matrix elements in the dimer basis. The resulting 
inter-dimer exchange terms can be expressed most explicitly in terms of 
transfer operators within the basis of two-dimer states. As an example, the 
off-diagonal components of $\widetilde{H}$ contributed by the $TD$-terms are 
given by 
\begin{eqnarray*}
\widetilde{H}^\mathrm{TD,off}_{dd'} & = & - {\textstyle \frac12} |J - J_2| \bigl[
\transfer{+S}{+0}  +  \transfer{+0}{+S} \\
& & \qquad\qquad\qquad + \transfer{+-}{0S} + \transfer{0+}{+S} \\
& & \qquad\qquad\qquad + \transfer{+S}{0+} + \transfer{0S}{+-} \\
& & \qquad\qquad\qquad + (+ \leftrightarrow -)  \bigr ],
\end{eqnarray*}
where the notation $| S+ \rangle$ denotes $|S\rangle_d \otimes | +\rangle_{d'}$
and the form is readily obtained with the help of Table \ref{tab:matele}.
The full sign-problem-free Hamiltonian in the dimer basis is the sum of 
the diagonal part, $H_d$, and the contributions from all such off-diagonal 
inter-dimer terms with their signs set to be non-positive. 

\begin{figure}[t!]
\centering
\includegraphics[width=0.95\columnwidth]{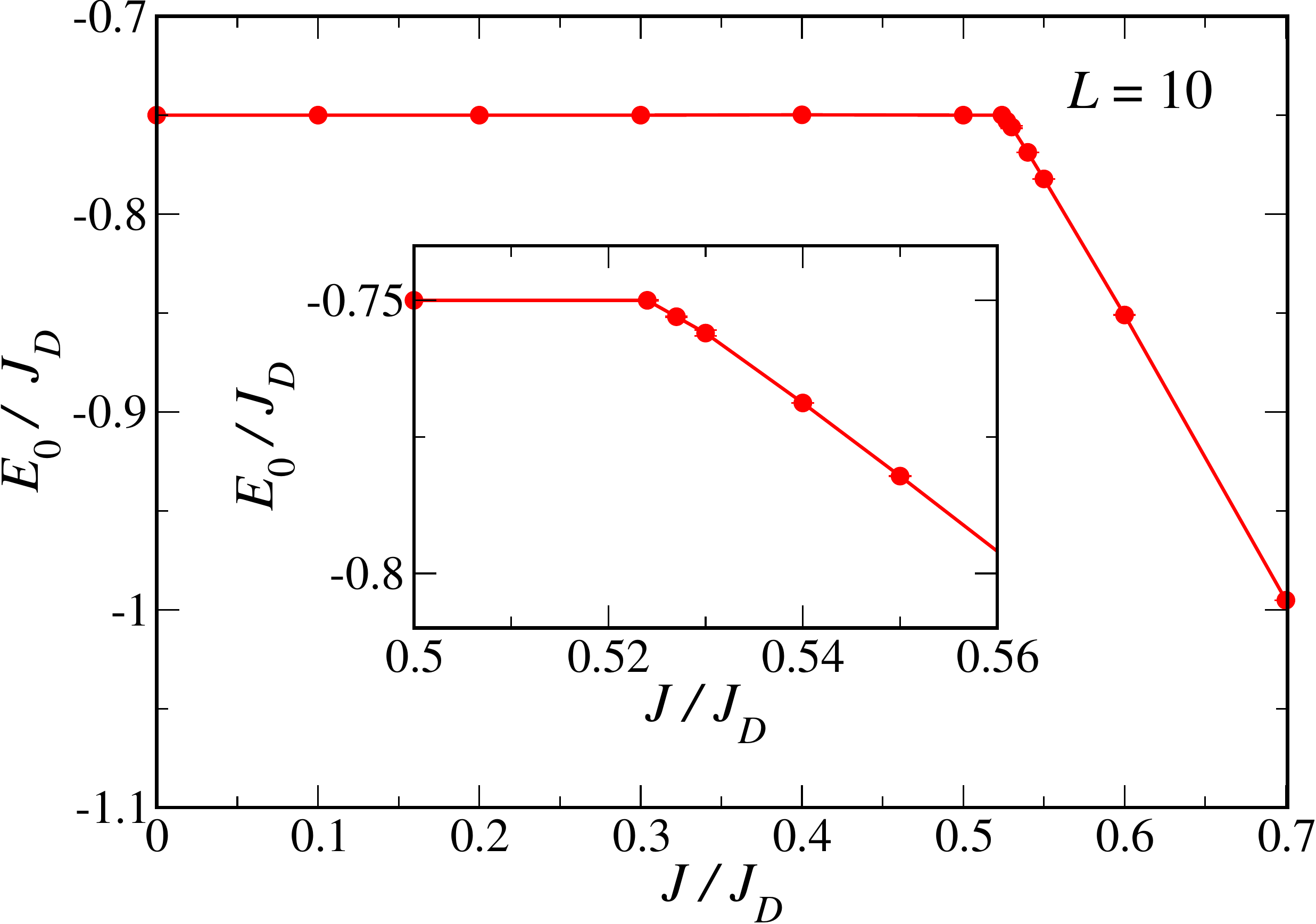}
\caption{Ground-state energy per dimer of the sign-problem-free model, 
$\widetilde{H}$, extracted from QMC simulations. The kink at $J/J_D = 0.526(1)$ 
signals a level-crossing out of the dimer singlet-product state.}
\label{fig:energy_bosonic_model}
\end{figure}

Because this Hamiltonian has no minus-sign problem by construction, it 
can be studied down to very low temperatures by QMC. The ground-state 
energy per dimer, $E_0$, for $\widetilde{H}$ is shown in 
Fig.~\ref{fig:energy_bosonic_model}. As expected, the ground state for 
weak inter-dimer coupling is the product of singlets on all dimer bonds, 
of course with energy $-{\textstyle \frac{3}{4}} J_D$ per dimer. As in 
the Shastry-Sutherland model, this product state remains as the exact 
ground state up to a fixed, finite value of the inter-dimer coupling, 
which we find to be $J = 0.526(1)J_D$ when $J_2 = 0$. At that coupling 
ratio, a level crossing takes place, signaling a first-order transition 
to another phase. Because this model is not physical, but useful only to 
discuss QMC simulations of the extended Shastry-Sutherland model, we have 
not tried to understand the precise nature of the high-$J$ phase of 
$\widetilde{H}$. We note only that it appears to extend up to very large 
$J/J_D$ with no sign of a further transition, but given the complicated 
form of the model we do not speculate on the physics in this regime.

By contrast, at small inter-dimer coupling it is straightforward to 
convince oneself that the ground state of the sign-problem-free 
Hamiltonian must be the same as that of the Shastry-Sutherland model. 
First, we observe that the singlet-product state is an element of the 
dimer basis in which the Hamiltonian is formulated. Thus the fact that it 
is an eigenstate of the Shastry-Sutherland model implies that all 
off-diagonal matrix elements involving that state must vanish. Because 
passing from the Shastry-Sutherland model to the sign-problem-free 
Hamiltonian involves only changing the signs of the positive off-diagonal 
matrix elements, all off-diagonal matrix elements involving the 
singlet-product state will still vanish in $\widetilde{H}$, implying that 
this state is an eigenstate of that Hamiltonian. Second, this state is 
clearly the ground state of the model with vanishing off-diagonal matrix 
elements, and it is separated from the first excited state by an energy 
equal to the intra-dimer coupling. A simple perturbative argument 
therefore implies that this situation has to remain true over a finite 
regime of parameter space where the off-diagonal matrix elements are 
small compared to the intra-dimer coupling.

Closing this section with a brief technical summary, we perform 
stochastic series expansion \cite{sse0} QMC simulations in the dimer 
basis \cite{rus1,Alet16} with directed loop updates \cite{sse1,sse2} to 
compute the thermodynamic properties of the Shastry-Sutherland model 
(\ref{ess}) and to characterize the sign problem in the extended model 
(\ref{extended}). These simulations perform an unrestricted sampling of 
the configuration space, meaning one not constrained to any subset of the 
Hilbert space defined by the $S^z$ and $D^z$ operators of the total system. 
We deploy a parallel tempering approach~\cite{rus1} to enhance state mixing, 
which is particularly important near the limit of the fully frustrated bilayer.
We access system sizes $N =$ 2$\times$$L$$\times$$L$ up to $L = 10$ and 
temperatures as low as $T = 0.01 J_D$ where the sign problem is mild. Where 
the sign problem is severe, we have worked down to average-sign values 
$\langle {\rm sign} \rangle' = 0.06$, where it is necessary to compensate by 
increasing the QMC sampling (the CPU time) by a factor of nearly 300.

\section{The minus sign}
\label{The minus sign}

Turning now to the minus sign in the model of Eq.~(\ref{extended}), it is 
always possible to simulate a model with a sign problem using QMC, by taking 
the absolute values of the weights, $|W_c|$, of each configuration $c$ from 
the corresponding sign-problem-free model. In this procedure, the average of 
any observable, $A$, is the ratio of the averages of the observable and of 
the sign \cite{EvertzLoop03,PhysRevLett.94.170201},
\begin{equation}
\label{eq:sign}
\langle \! A \rangle = \frac{\sum_c W_c A_c}{\sum_c W_c} = \frac{\sum_c 
\text{sign}(W_c) |W_c| A_c}{\sum_c \text{sign}(W_c) |W_c|} = \frac{\langle 
\text{sign} A \rangle'}{\langle \text{sign} \rangle'}. 
\end{equation}
Here the notation $\langle X \rangle'$ indicates that $|W_c|$ is obtained from 
the sign-problem-free Hamiltonian, in which the weights are readily sampled, 
but we stress that the physics of the original model is contained in the 
signs, $\text{sign}(W_c) = W_c/|W_c|$, of every configuration $c$, which appear 
in both the numerator and the denominator of $\langle A \rangle$. In a typical 
frustrated quantum spin model, this approach can no longer be used when the 
temperature becomes low compared to the energy scales set by the coupling 
strengths, because then the average sign, $\langle \text{sign} \rangle'$ in 
the denominator of Eq.~(\ref{eq:sign}), tends to zero, inducing error bars 
larger than the signal. 

\begin{figure}[t!]
\includegraphics[width=0.95\columnwidth]{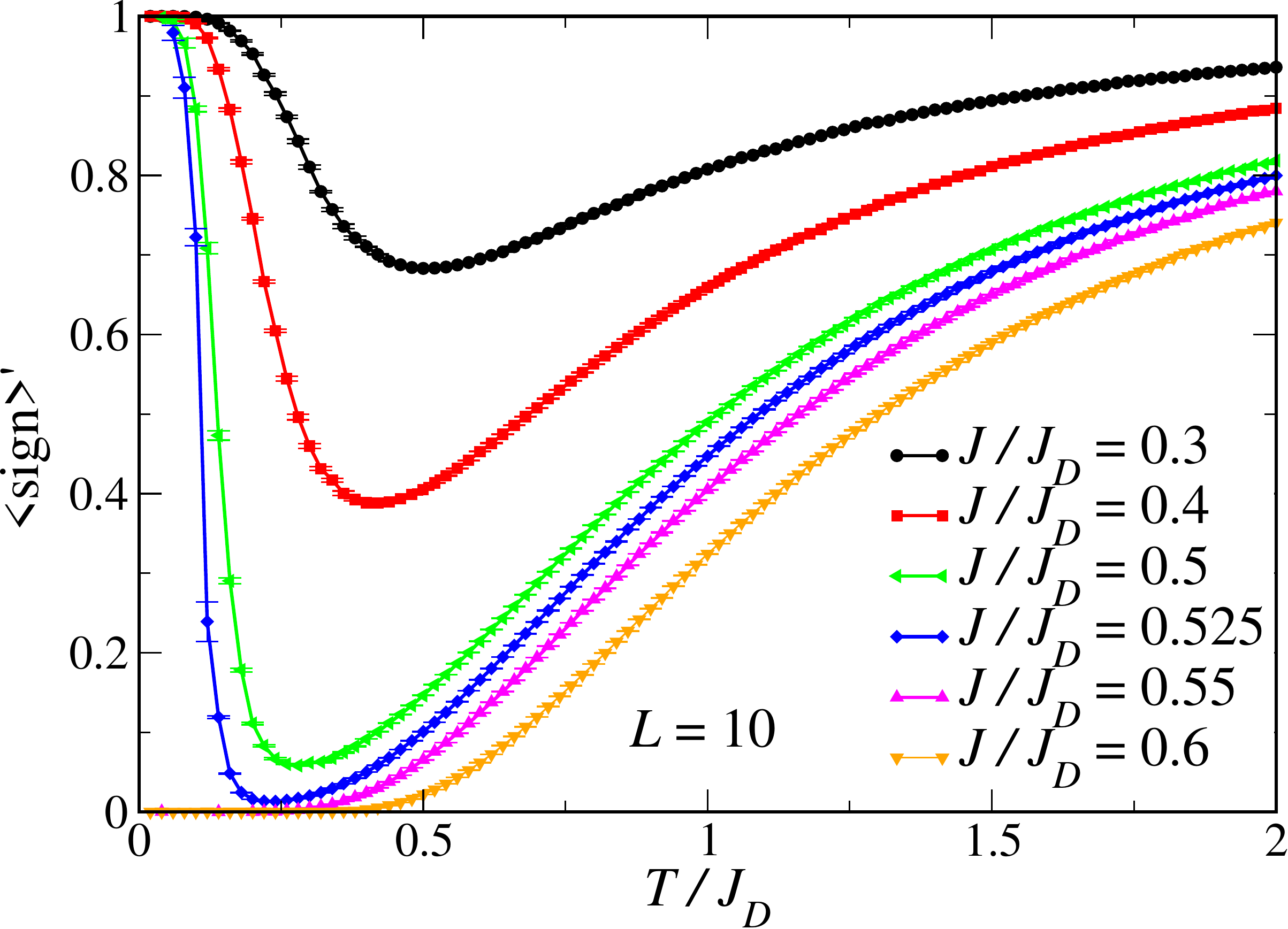}
\caption{Temperature dependence of the average sign, $\langle \text{sign} 
\rangle'$, computed for the Shastry-Sutherland model [Eq.~(\ref{ess})] with 
different values of the coupling ratio.}
\label{fig:minus_sign}
\end{figure}

The central result of the present contribution is reported in 
Fig.~\ref{fig:minus_sign}. While $\langle \text{sign} \rangle'$ for the 
Shastry-Sutherland model does indeed become small at temperatures below 
$J_D$, it increases again at low temperatures and recovers to a value of 
precisely 1 at zero temperature. This behavior occurs provided that the 
ground state of the sign-problem-free model is the singlet-product state, 
and thus it holds up to the coupling ratio $J/J_D = 0.526(1)$. Above that 
coupling value, the behavior of $\langle \text{sign} \rangle'$ is typical 
of any general model with a minus sign, i.e.~the average becomes very 
small and never increases again 
\cite{EvertzLoop03,PhysRevLett.94.170201}.

The fact that the average sign goes rigorously to 1 at zero temperature 
is a simple consequence of the fact that both the Shastry-Sutherland 
model and its sign-problem-free counterpart have the same ground state. 
Then the denominator of Eq.~(\ref{eq:sign}) is strictly equal to 1 and 
the average of any quantity is its ground-state expectation value. This 
should be contrasted with the frustrated ladder away from perfect 
frustration, where the ground state cannot be expressed exactly in the 
dimer basis and periodic boundary conditions introduce components with a 
minus sign \cite{rus3}. In that case, the average sign also increases 
again as the temperature is lowered, but recovers only to a value close, 
i.e.~not exactly equal, to 1.

\begin{figure}[t!]
\includegraphics[width=\columnwidth]{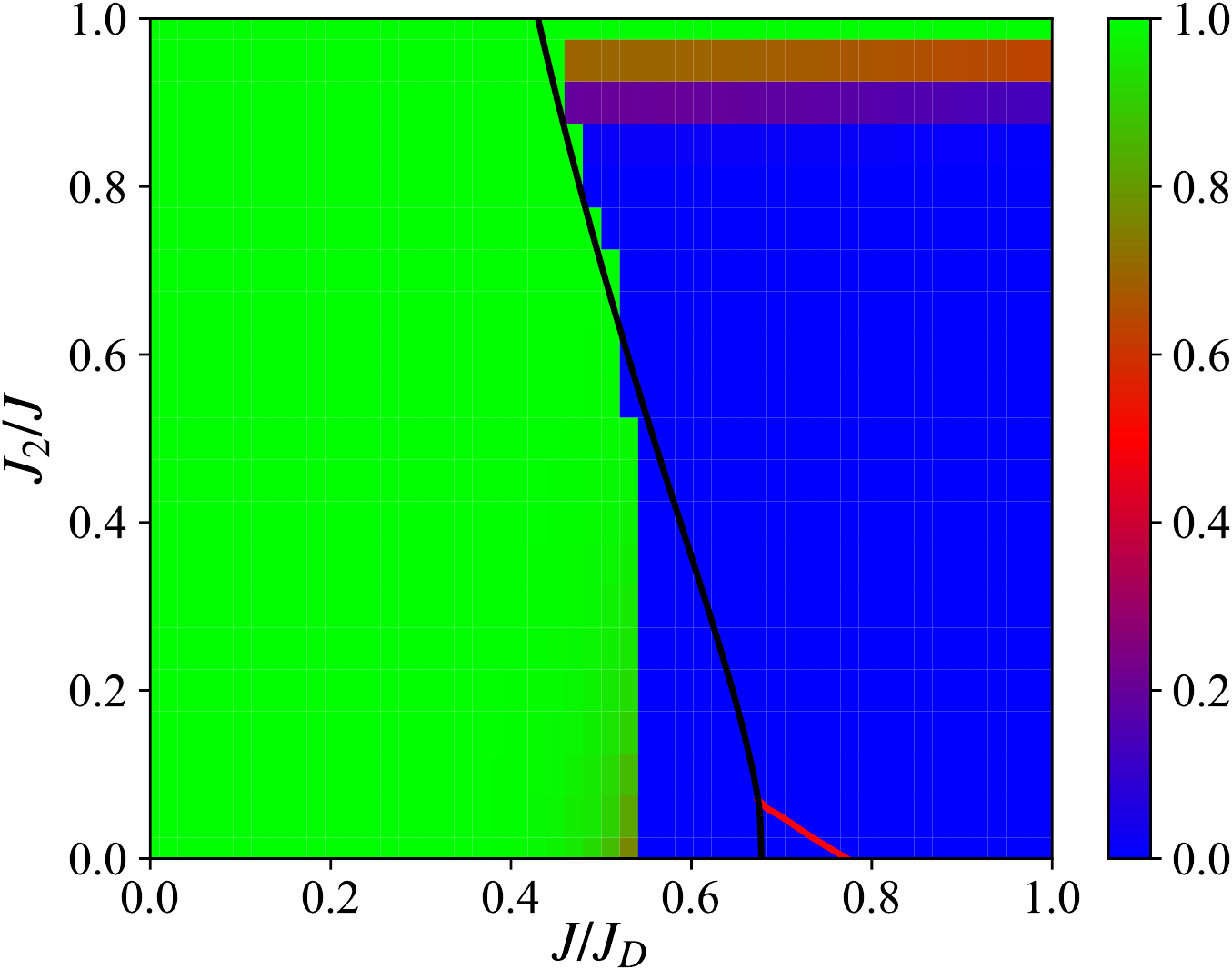}
\caption{Average sign, $\langle \text{sign} \rangle'$, computed at a 
temperature $T = 0.1 J_D$ throughout the phase diagram of the extended 
Shastry-Sutherland model [Eq.~(\ref{extended})] for a system of 10$\times$10 
dimers. Solid lines reproduce the phase boundaries computed by iPEPS and 
shown in Fig.~\ref{fig:phase_diagram_peps}.} 
\label{fig:phase_diagram_sign}
\end{figure}

\begin{figure}[t]
\includegraphics[width=0.95\columnwidth]{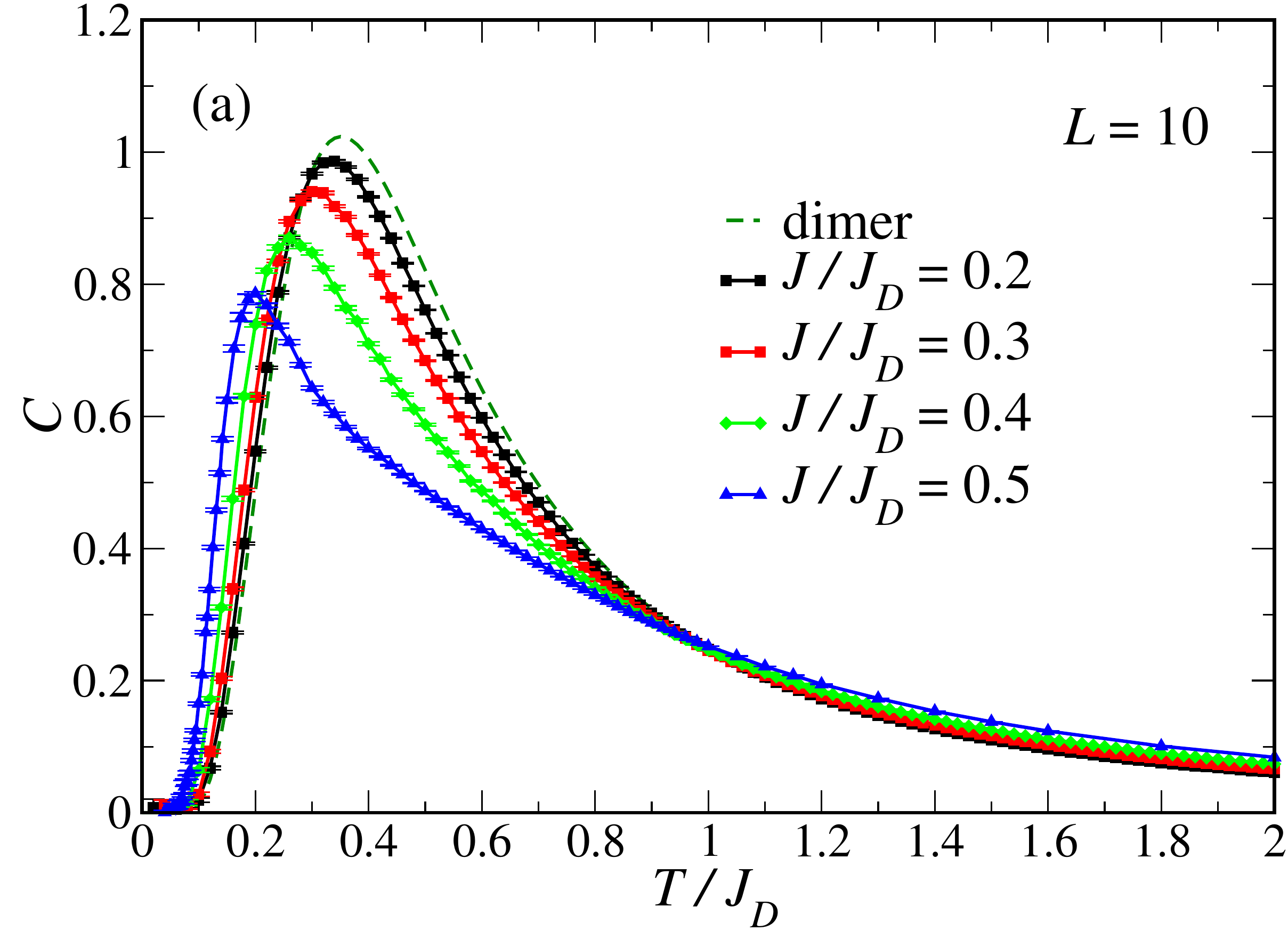}
\includegraphics[width=0.95\columnwidth]{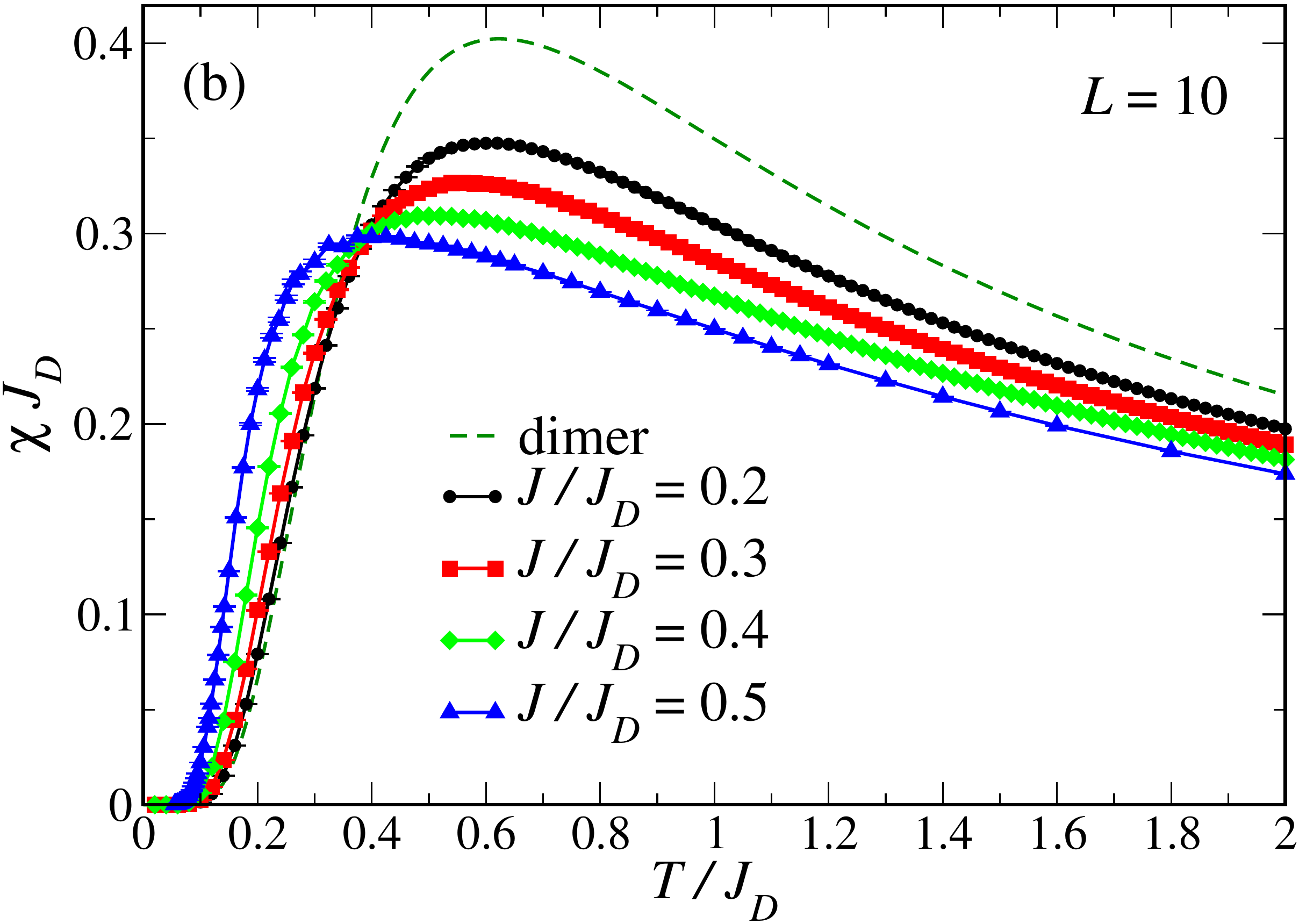}
\caption{ (a) Magnetic specific heat, $C(T)$, and (b) susceptibility, 
$\chi(T)$, of the Shastry-Sutherland model computed by QMC for systems of 
size $L = 10$, shown per dimer for different values of the coupling ratio.}
\label{fig:cchi}
\end{figure}

Our motivation for considering the extended Shastry-Sutherland model of 
Eq.~(\ref{extended}) was that the limit ($J_2 = J$) of the fully 
frustrated bilayer is completely sign-problem-free. One may therefore 
hope that a significant fraction of the phase diagram of 
Fig.~\ref{fig:phase_diagram_peps}, in the regime around this limit, may 
have only a mild sign problem and would thus be amenable to QMC. To 
investigate this possibility, we have calculated $\langle \text{sign} 
\rangle'$ for the extended model by working on a system of fixed size $L 
= 10$ and at a fixed temperature $T = 0.1J_D$. As 
Fig.~\ref{fig:phase_diagram_sign} makes clear, the average sign is 
essentially equal to 1 in a large portion of the singlet-product phase. 
The border of the sign-problem-free region is almost vertical near the 
Shastry-Sutherland limit (small $J_2$), which is a consequence of the 
phase transition at $J/J_D = 0.526$ in the (unphysical) sign-problem-free 
model, as discussed in Sec.~\ref{The models} and 
Fig.~\ref{fig:energy_bosonic_model}. For $J_2$ values beyond 
approximately $0.5\,J$, the boundary of the sign-problem-free region 
matches quite accurately the physical boundary to the antiferromagnetic 
phase, which we show in Fig.~\ref{fig:phase_diagram_sign} by reproducing 
the transition line from the ground-state phase diagram computed by iPEPS 
(Fig.~\ref{fig:phase_diagram_peps}). In the fully frustrated limit, $\langle 
\text{sign} \rangle'$ exhibits no transition, which is to be expected 
because the physical model is completely free of any sign problems here 
\cite{Alet16,NgYang17,rus4}. However, the sign problem manifestly grows 
very rapidly with ``detuning'' ($J_2 \ne J$) away from the fully frustrated 
line, leaving very little additional parameter space where one might hope 
to use QMC to study, for example, the dimerized-to-antiferromagnetic phase 
transition. We comment that, in the regime of a dominant interaction $J$, 
where the ground state of the Shastry-Sutherland model (\ref{ess}) is 
antiferromagnetically ordered (right side of 
Fig.~\ref{fig:phase_diagram_peps}), one may perform sign-problem-free 
QMC simulations in the standard basis of spin configurations only for 
$J_D = 0$. For any finite values $J_D > 0$, these simulations are again 
plagued by a severe sign problem, which prevents us from examining the 
transition regime out of the antiferromagnetic phase.


\begin{figure*}[p]
\includegraphics[height=0.4\textwidth]{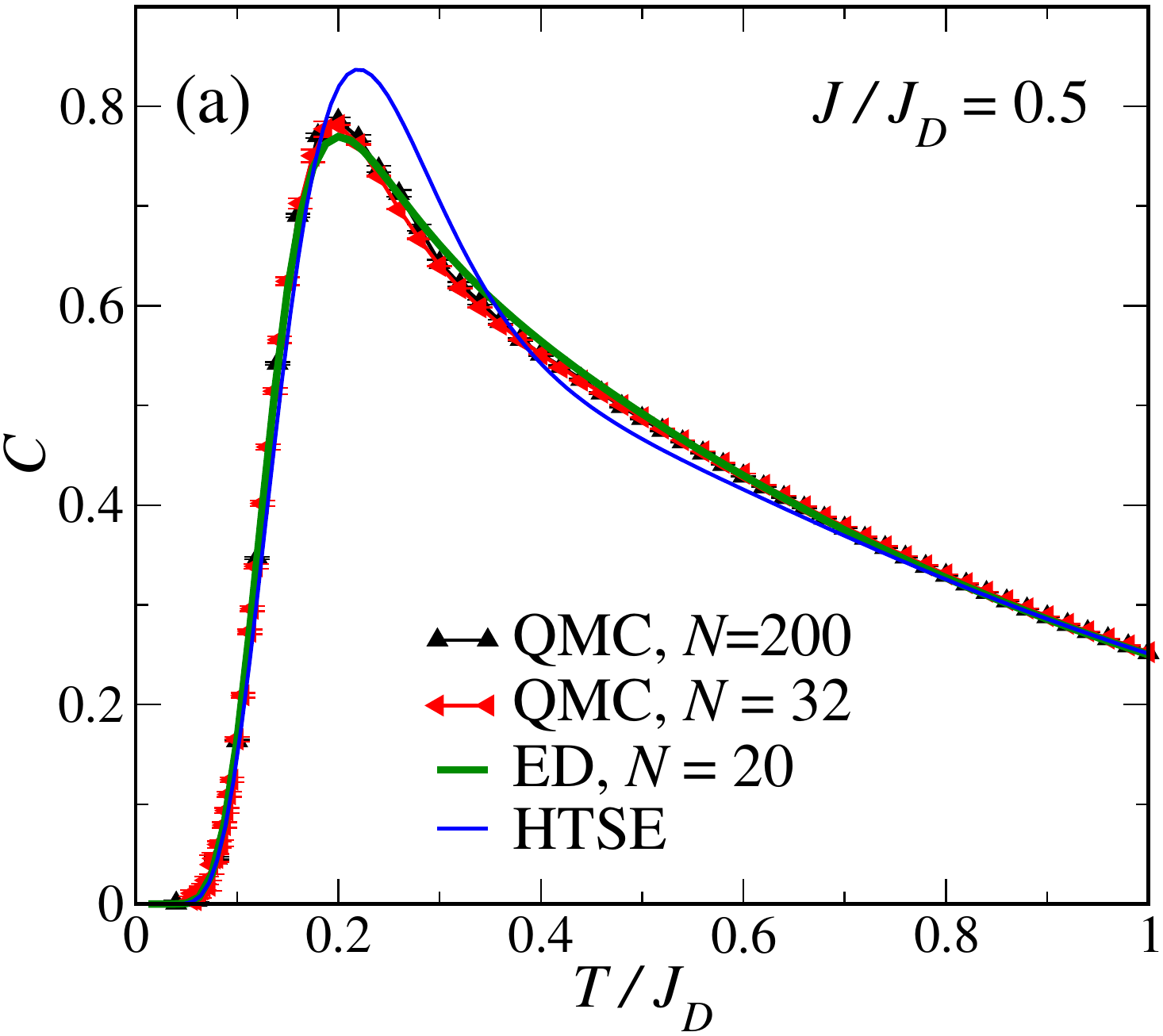}\hfill%
\includegraphics[height=0.4\textwidth]{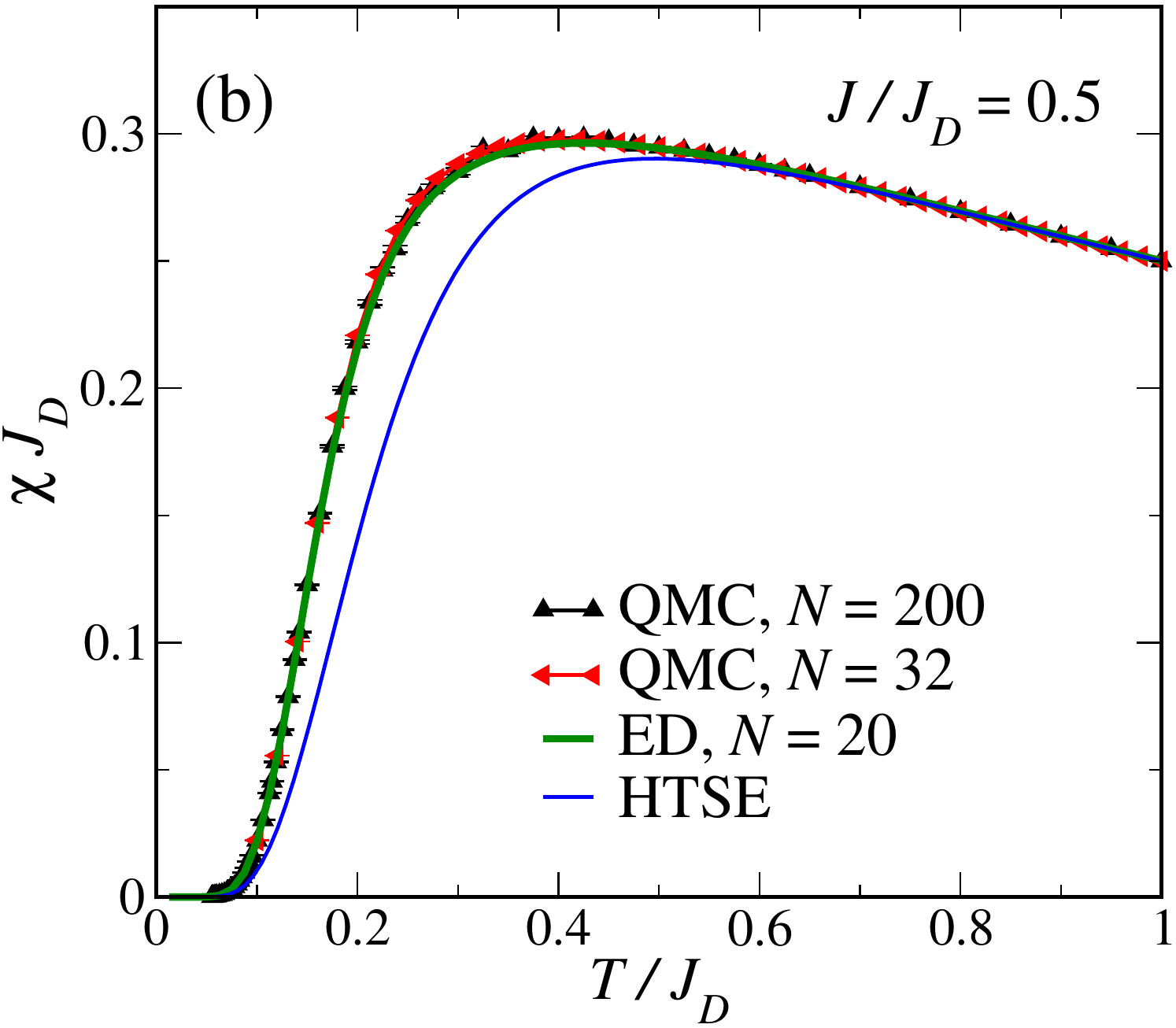}\\
\includegraphics[height=0.4\textwidth]{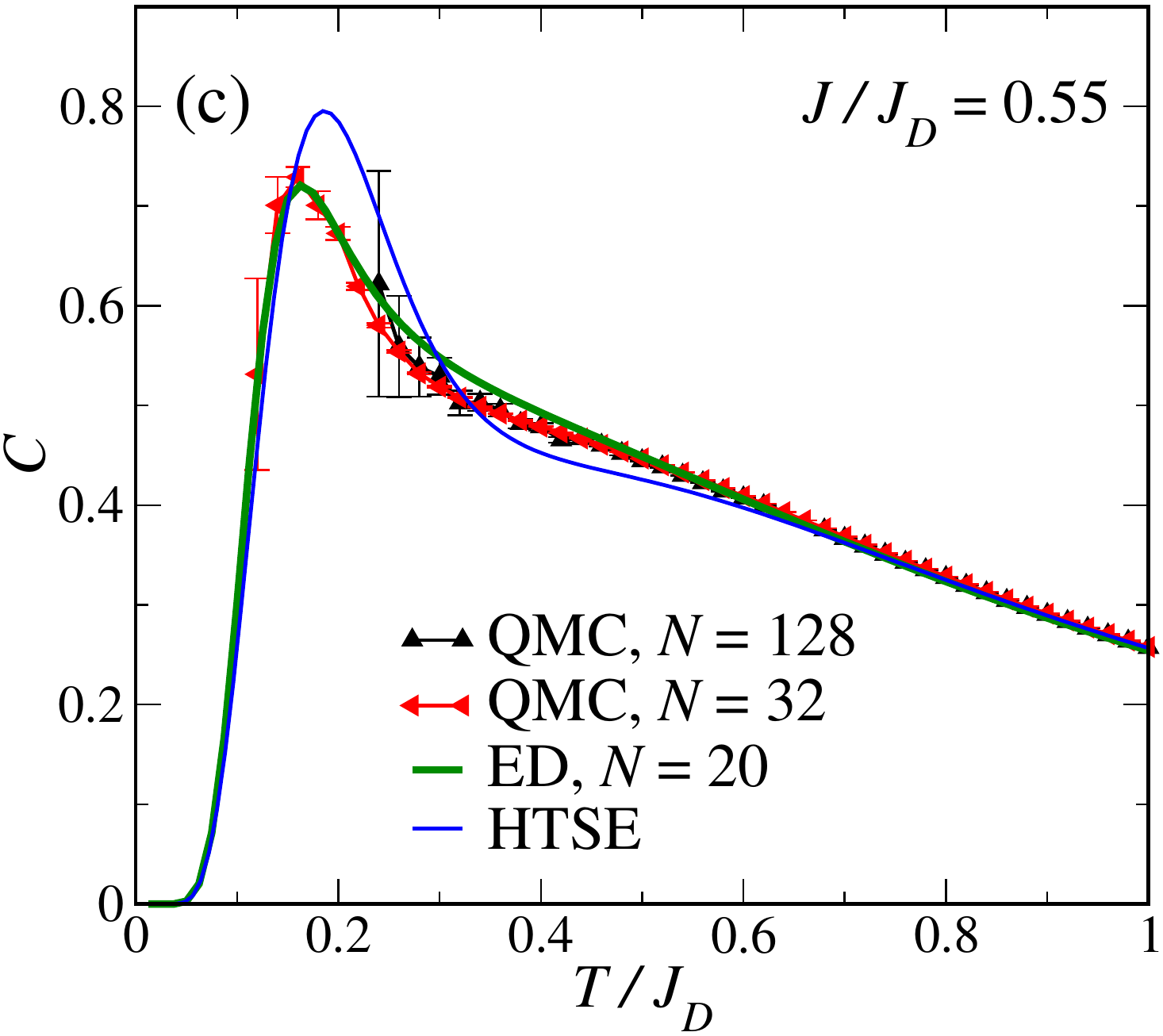}\hfill%
\includegraphics[height=0.4\textwidth]{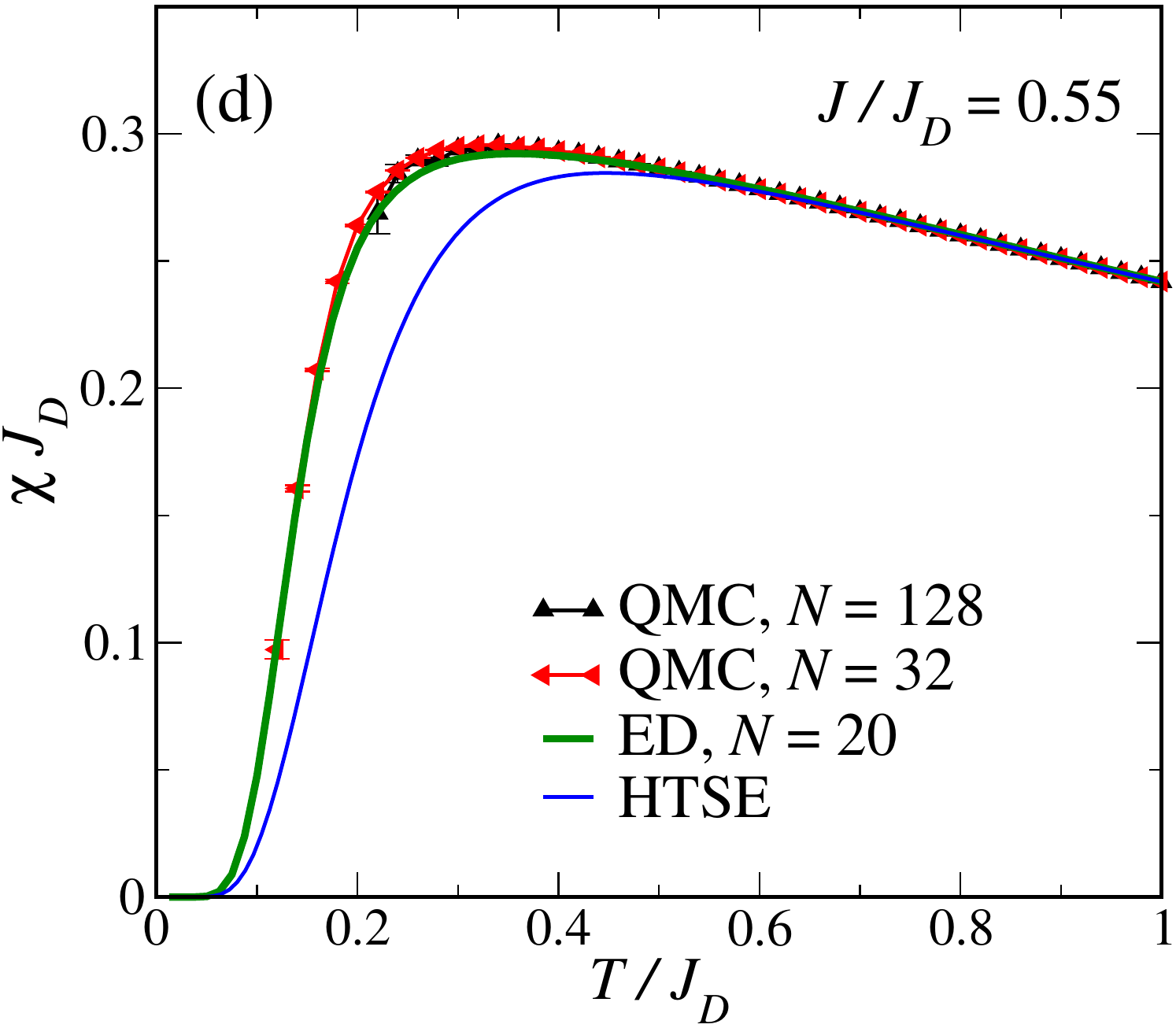}\\
\includegraphics[height=0.4\textwidth]{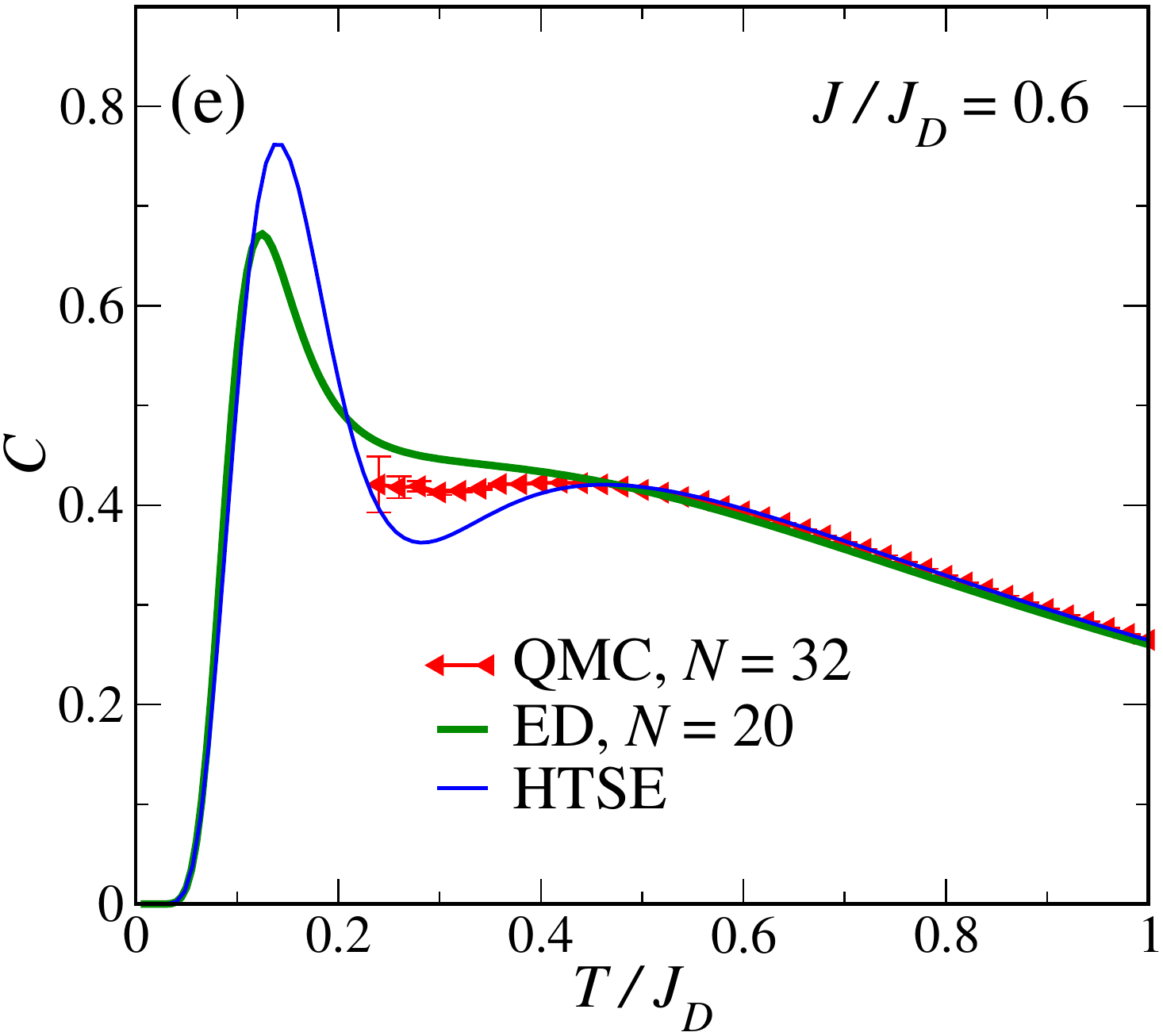}\hfill%
\includegraphics[height=0.4\textwidth]{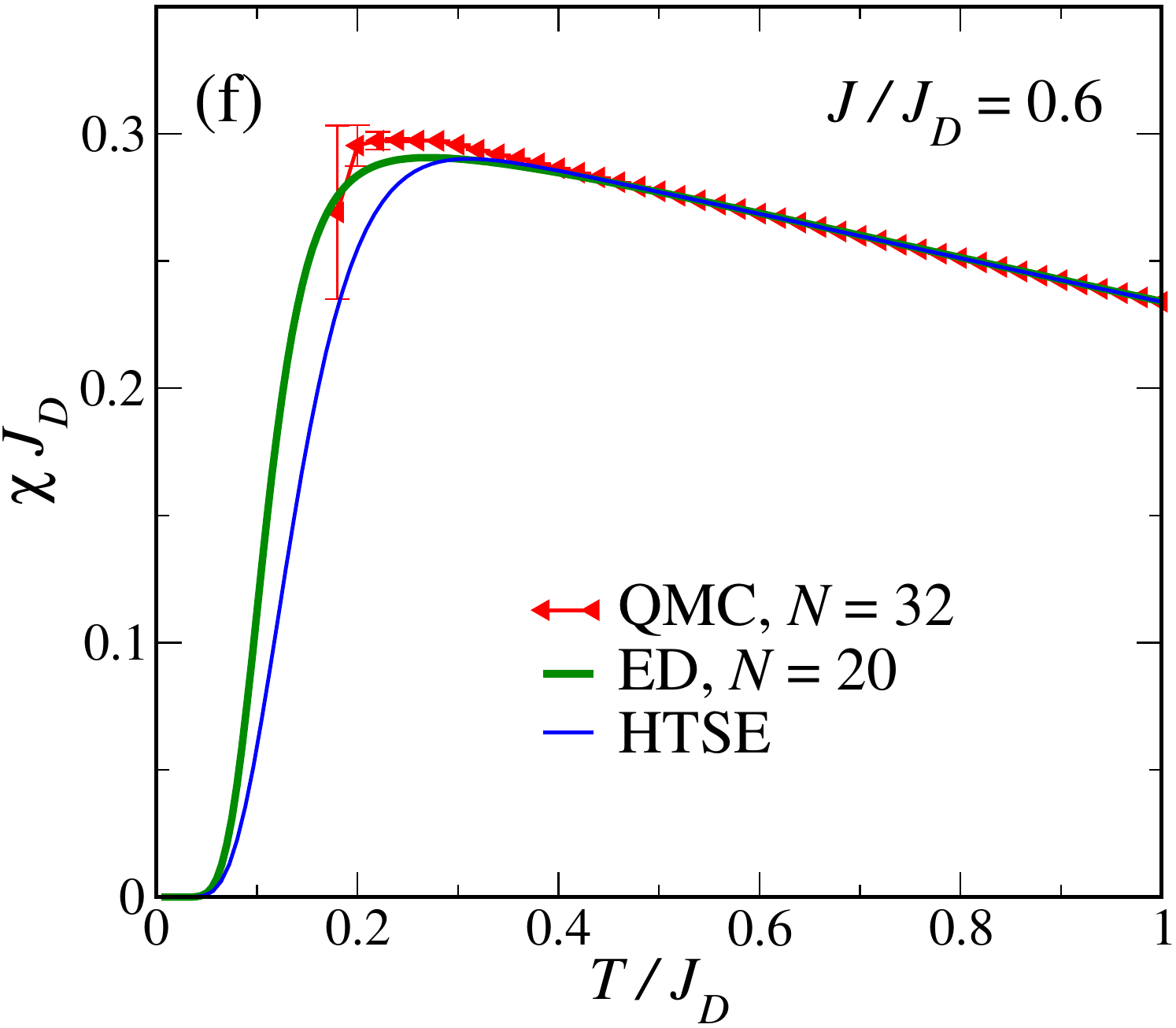}
\caption{(a,c,e) Magnetic specific heat, $C(T)$, and (b,d,f) susceptibility, 
$\chi(T)$, of the Shastry-Sutherland model computed by QMC and shown per dimer 
for the largest system sizes feasible at coupling ratios $J/J_D = 0.5$ (a,b), 
0.55 (c,d), and 0.6 (e,f).}
\label{fig:Cchi}
\end{figure*}


\section{Thermodynamic Calculations}
\label{Physical results}

For our calculations of thermodynamic properties we focus on the original 
Shastry-Sutherland model of Eq.~(\ref{ess}), i.e.~the case $J_2 = 0$ in 
Eq.~(\ref{extended}). By inspection of Fig.~\ref{fig:minus_sign}, the 
average sign for a coupling ratio such as $J/J_D = 0.5$ falls (in 
calculations using $L = 10$) to values as low as $0.06$ over a 
significant range of intermediate temperatures. Nonetheless, as a 
consequence of our observations concerning the ground state 
(Sec.~\ref{The models}) and the minus sign (Sec.~\ref{The minus sign}), 
it remains possible to obtain very accurate results in the regime $J 
\lesssim 0.5\,J_D$ for the magnetic specific heat, $C(T)$, and 
susceptibility, $\chi(T)$, which are shown respectively in 
Figs.~\ref{fig:cchi}(a) and \ref{fig:cchi}(b).

Figure \ref{fig:cchi} shows data obtained by simulations for clusters of 
10$\times$10 dimers, corresponding to a system containing $N = 200$ $S = 
1/2$ spins. In this regime, finite-size effects are sufficiently small 
that these results can be considered as fully representative of the 
thermodynamic limit. For this reason, we have not performed simulations 
for still larger values of $N$, although this would be completely feasible due 
to the rather mild sign problem in this parameter regime. In the limit $J = 0$, 
we recover the result for decoupled dimers, which is known analytically 
\cite{Bleaney451,PhysRevB.61.9558,PhysRevB.74.174421,rus1} and is represented 
by the dashed lines. As the ratio $J/J_D$ is increased, $\chi(T)$ shows a 
flattening of its maximum accompanied by a downward shift of its 
low-temperature flank [Fig.~\ref{fig:cchi}(b)], indicating a decreasing 
spin gap. $C(T)$ exhibits a similar suppression of both spin gap and peak 
position [Fig.~\ref{fig:cchi}(a)]; although the full response remains broad 
in temperature, there is a distinct sharpening of the low-temperature peak 
as $J/J_D$ approaches 0.5.

In Fig.~\ref{fig:Cchi} we study the challenging regime of coupling ratios 
between $J/J_D = 0.5$ and the transition from dimer to plaquette order. 
This is also the region of interest to experiment, for the description of 
SrCu$_2$(BO$_3$)$_2$. In addition to QMC data, here we also show ED results, 
obtained by full diagonalization of the relevant Hamiltonians for clusters 
of $N = 20$ spins, and the results of interpolated high-temperature series 
expansions (HTSEs); technical details of the HTSE approach may be found 
in App.~\ref{sec:appHighTser}. Figures \ref{fig:Cchi}(a) and 
\ref{fig:Cchi}(b) revisit $C(T)$ and $\chi(T)$ for the case $J/J_D = 0.5$ 
in order to compare our $N = 200$ QMC data (Fig.~\ref{fig:cchi}) with 
results for $N = 32$. The negligible deviations between the two data sets 
confirm that $N = 200$ is indeed well in the thermodynamic limit (whence, 
again, we did not perform simulations for any larger $N$, although this 
would still be possible at $J/J_D = 0.5$). However, minor deviations from 
the $N = 20$ ED data do start to become visible around the maximum of the 
specific heat, indicating the onset of finite-size effects for $N \le 20$ 
at $J \ge 0.5\,J_D$.

Turning to our HTSE calculations, the interpolated tenth-order HTSEs 
capture the qualitative behavior visible in the QMC and ED data for 
$J/J_D = 0.5$ and improve upon previous seventh-order studies 
\cite{PhysRevB.60.6608}, most notably in that the interpolation scheme 
outlined in App.~\ref{sec:appHighTser} enhances the stability of the 
expansion in comparison to earlier work. However, in contrast to the 
situation at smaller values of $J/J_D$ (App.~\ref{sec:appHighTser}), our 
HTSEs are not able to reproduce the QMC and ED results for $J/J_D = 0.5$ 
with quantitative accuracy. With a view to understanding the limits of 
the present procedure, we note that the low-temperature edge of $C(T)$, 
which is normally controlled by the spin gap, is reproduced very well in 
Fig.~\ref{fig:Cchi}(a), whereas this is not the case for $\chi(T)$ in 
Fig.~\ref{fig:Cchi}(b). Technically, a possible reason why $C(T)$ is 
relatively better behaved may lie in the additional energy and entropy 
sum rules that can be used to stabilize the interpolation 
\cite{PhysRevB.63.134409,PhysRevLett.114.057201,PhysRevE.95.042110}. 
Physically, one may suspect this discrepancy of indicating the onset of a 
regime where the low-temperature thermodynamics are no longer controlled 
in a conventional way by a small number of low-lying excited states 
\cite{rus1}, and we return to this point below.

At $J/J_D = 0.55$ and $0.6$, the average sign in the Shastry-Sutherland 
model no longer recovers to 1 at low temperatures 
[Fig.~\ref{fig:minus_sign}]. Unsurprisingly, dimer-basis QMC simulations 
become very much more challenging in this regime and we are forced to 
reduce the system size in order to reach meaningfully low temperatures. 
System sizes of $N = 32$ are required to reach temperatures below the 
maximum of the specific heat at $J = 0.55\, J_D$ 
[Fig.~\ref{fig:Cchi}(c)], but comparison with $N = 128$ data does indicate 
that $N = 32$ remains sufficient to keep deviations from the 
thermodynamic limit within the statistical error bars. By contrast, ED 
results for $N = 20$ at $J = 0.55\,J_D$ show definite finite-size 
effects, specifically in the region $0.2 < T/J_D < 0.4$ in $C(T)$ and 
around the maximum of $\chi(T)$. The coupling ratio $J/J_D = 0.6$, shown 
in Figs.~\ref{fig:Cchi}(e) and \ref{fig:Cchi}(f), marks the outer limit 
of the regime where the low-temperature behavior of the 
Shastry-Sutherland model can be considered to be under control in any 
quantitative sense. Comparison between $N = 20$ ED and $N = 32$ QMC data 
shows that $C(T)$ [Fig.~\ref{fig:Cchi}(e)] remains subject to very 
significant finite-size effects for $T/J_D \lesssim 0.5$, where it is 
possible that non-monotonic behavior sets in, while it is difficult to 
benchmark anything below the maximum of $\chi(T)$ 
[Fig.~\ref{fig:Cchi}(f)].

\begin{figure}[t!]
\centering
\includegraphics[width=\columnwidth]{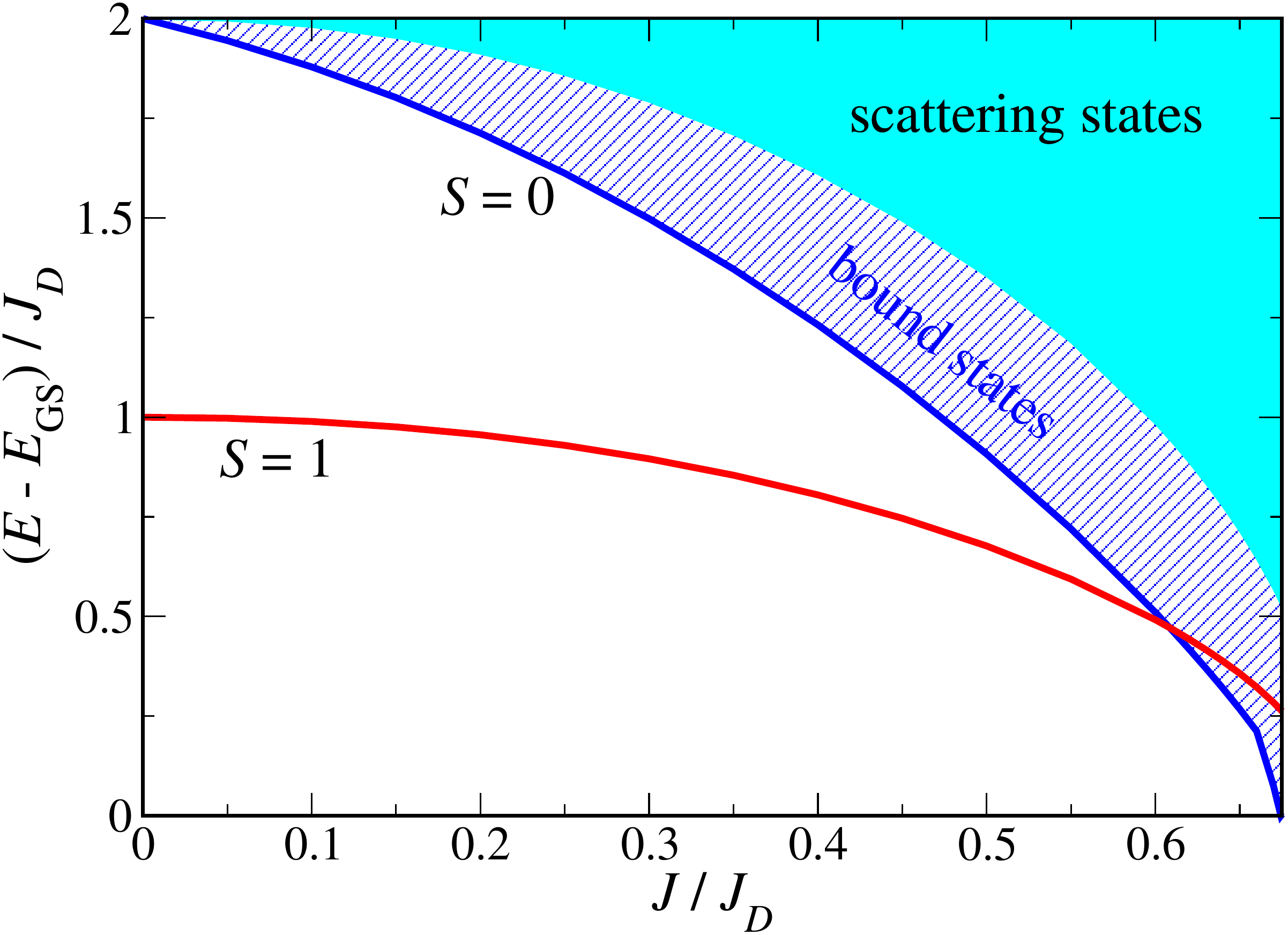}
\caption{Schematic representation of the excitation spectrum of the 
Shastry-Sutherland model shown as a function of coupling ratio, based on 
ED calculations using a cluster of $N = 36$ spins. Energies are measured 
with respect to the ground-state energy, $E_{\rm GS} = - \frac{3}{8} N J_D$. 
Solid lines denote the gaps to the lowest triplet ($S = 1$, red) and 
singlet ($S = 0$, blue) excited states. The band width of the one-particle 
triplet excitations is approximately the thickness of the red line. The 
hatched region represents continua of dispersive $S = 0$ and $S = 1$ 
two-particle bound states. The shaded region, whose lower boundary is 
given by twice the triplet gap, indicates the regime where many-particle 
scattering continua are allowed.}
\label{fig:Gap36}
\end{figure}

The results of Fig.~\ref{fig:Cchi} confirm the physical trends observed in 
Fig.~\ref{fig:cchi}, namely the downward shift of the low-temperature rise 
in both $C(T)$ and $\chi(T)$ with increasing $J/J_D$, accompanied by a 
flattening of the maximum in $\chi(T)$ and a sharpening of the peak in 
$C(T)$. The emergence of this distinctive maximum at a temperature scale 
very low in comparison with the coupling constants constitutes the 
dominant thermodynamic feature as one approaches the first-order transition 
from the dimer-singlet to the plaquette phase at $J/J_D \approx 0.675$ 
\cite{CorbozMila13}. This behavior is analogous to that observed on 
approaching the boundary of the rung-singlet phase in highly frustrated spin 
ladders \cite{rus1,rus3}, where its origin was traced to the presence of many 
low-lying bound rung-triplet excitations. Our results suggest that the same 
type of bound-state mechanism is at work in the less constrained 2D system, 
and that the emergence of the low-temperature maximum in the specific heat 
is its clearest thermodynamic fingerprint.

To expand upon this point, in Fig.~\ref{fig:Gap36} we show a schematic 
representation of the excitation spectrum of the Shastry-Sutherland model 
in the thermodynamic limit as a function of $J/J_D$. The solid red and blue 
lines mark respectively the gaps to the lowest-lying triplet and singlet 
states, which we have extracted from earlier $N = 36$ ED calculations 
\cite{PhysRevLett.111.137204}; the former have their origin in single 
dimer-triplet excitations, which are only very weakly dispersive, and the 
latter in bound pairs of dimer triplets. We note that the decrease in energy 
not only of the spin gap but also of the two-particle bound states on 
increasing $J/J_D$ is already well documented \cite{PhysRevLett.85.3958,
Fukomoto00,PhysRevLett.86.520,MiUeda03}. A considerable number of dispersive 
singlet and triplet bound states remains below the edge of the two-particle 
continuum \cite{PhysRevLett.85.3958}, as represented by the blue hatched 
region in Fig.~\ref{fig:Gap36}. We draw attention in particular to the fact 
that the gap of the lowest singlet mode decreases faster with coupling ratio 
than the triplet gap, until the two cross at $J/J_D \approx 0.61$ on the 
$N = 36$ cluster. While the singlet spectrum remains unknown in detail, it 
is likely that these bound states are responsible for the sharpening peak in 
$C(T)$ at $J/J_D \ge 0.6$. The fact that the gap of the lowest singlet bound 
state for $N = 36$ closes very near the boundary of the dimer singlet-product 
phase, $J/J_D \approx 0.675$ \cite{CorbozMila13} (right border of 
Fig.~\ref{fig:Gap36}), is expected to indicate the crossing of levels 
occurring at the first-order transition to the plaquette phase. 

A more detailed analysis of the evolution with $J/J_D$ of the $n$-particle 
bound states in the ED spectrum with $n > 2$ is an involved problem that 
we defer to a future study. We stress that, over most of the singlet-product 
regime of the phase diagram, and certainly the range $J/J_D \le 0.5$, the 
thermodynamic response of the Shastry-Sutherland model (Fig.~\ref{fig:cchi}) 
should be characteristic of just one gap, that to the lowest triplet. Only 
beyond this region, coincidentally in the zone where QMC becomes dramatically 
more difficult [Figs.~\ref{fig:Cchi}(c)-\ref{fig:Cchi}(f)], might the 
proximity of the lowest singlet state(s) indeed begin to play a role 
(Fig.~\ref{fig:Gap36}).

\section{Conclusions and Perspectives}
\label{Conclusions}

We have shown that, even for models where QMC simulations suffer from a 
minus-sign problem, it may be possible to obtain extremely accurate results 
for the low-temperature thermodynamics. A sufficient condition is that the 
ground states of the physical model and of the corresponding sign-problem-free 
model, constructed by making all off-diagonal matrix elements non-positive, 
be the same. This condition has allowed us to compute numerically exact results 
for the magnetic specific heat and susceptibility of the Shastry-Sutherland 
model throughout the parameter range where the ratio of the inter- to 
intra-dimer couplings is less than or equal to 0.526(1). 

This is the regime of coupling ratios where the ground state of both models 
is a product of singlets on every dimer bond, the state about which Shastry 
and Sutherland constructed their Hamiltonian. With regard to the material 
realizing the Shastry-Sutherland model, it is of course unfortunate that 
this critical ratio for the success of QMC is smaller than the coupling 
ratio in SrCu$_2$(BO$_3$)$_2$ \cite{PhysRevLett.82.3168}, which is believed 
to be approximately $0.63$ \cite{MiUeda03}. Because the real Shastry-Sutherland 
model has, at this coupling ratio, not yet undergone the phase transition to 
the plaquette state, we are investigating possible modifications to the 
conventional sign-problem-free model introduced in Sec.~\ref{The models} 
with a view towards making the weights sampled in this model applicable at 
coupling ratios larger than 0.526.

Our QMC results offer considerable perspective on other numerical approaches 
to the thermodynamics of the Shastry-Sutherland model. Clearly finite-size 
effects become increasingly important at $J/J_D > 0.5$ and thus ED studies, 
particularly using clusters of $N \le 20$ spins \cite{MiUeda99,MiUeda00,
MiUeda03}, must be interpreted with care at low temperatures and especially 
at $J/J_D \approx 0.63$. This highlights the importance of ED variants that 
access larger $N$ by avoiding full diagonalization, such as that applied 
recently \cite{SSR18} to compute the thermodynamic properties of a kagome 
cluster with $N = 42$ $S = 1/2$ spins. We have also used our QMC results to 
benchmark some recent advances in HTSE approaches. While this comparison 
demonstrates qualitative progress, in that the problem of low-temperature 
divergences, which plagued previous HTSE implementations \cite{MiUeda99,
PhysRevB.60.6608}, can be overcome by suitable interpolation schemes, it 
shows at the quantitative level that HTSE for the Shastry-Sutherland model 
remains limited by the maximum accessible expansion order of ten. Consequently, 
the accuracy of our HTSEs remains below that of QMC and even small-system ED 
over the full phase diagram of the model. A combination of deriving 
higher-order series (the 17th order has been attained in a recent study 
\cite{PhysRevLett.114.057201} of the kagome lattice) and more refined 
interpolation schemes \cite{PhysRevB.63.134409,PhysRevLett.114.057201,
PhysRevE.95.042110} may offer a competitive HTSE approach to the parameter 
regime relevant for SrCu$_2$(BO$_3$)$_2$.

Beyond the Shastry-Sutherland model, our results imply that QMC simulations 
should be possible for any frustrated model whose ground state is known 
exactly, provided that the Hamiltonian matrix can be expressed in a basis 
that contains this exact ground state. We anticipate that this observation 
will open up the field of QMC calculations of the thermodynamics for a range 
of frustrated quantum spin systems, most straightforwardly those constructed 
in order to possess exact dimer- and plaquette-product ground states. Here 
we have explored the extension of the Shastry-Sutherland model to the limit 
of the fully frustrated bilayer, where the sign problem is entirely absent, 
and demonstrate by comparison with iPEPS calculations of the ground-state 
phase diagram how the extent of the sign problem can be understood.

\begin{acknowledgments}

This work was supported by the Deutsche Forschungsgemeinschaft (DFG) under 
Grants FOR1807 and RTG 1995, by the Swiss National Science Foundation (SNF), 
and by the European Research Council (ERC) under the EU Horizon 2020 research 
and innovation programme (Grant No.~677061). We thank the IT Center at RWTH 
Aachen University and the JSC J\"ulich for access to computing time through 
JARA-HPC.

\end{acknowledgments}

\appendix

\section{iPEPS Calculations}
\label{sec:appipeps}

The ground-state phase diagram in Fig.~\ref{fig:phase_diagram_peps} was 
obtained by means of a variational tensor-network ansatz known as an infinite 
projected entangled pair state (iPEPS) \cite{tnref-verstraete04,tnref-nishio04,
tnref-jordan08}. An iPEPS consists of a unit cell of local tensors that is 
repeated over the lattice. Each local tensor has one physical index which, 
for the present model, represents the two sites on a dimer, and four 
auxiliary indices that connect neighboring local tensors to form a square 
geometry in accord with the lattice structure shown in Fig.~\ref{fig:model}(b). 
The auxiliary vector spaces have a dimension $D$, the bond dimension, which 
controls the accuracy of the ansatz, in that higher $D$ values allow more 
entanglement to be captured by the iPEPS. All three of the phases in 
Fig.~\ref{fig:phase_diagram_peps} can be represented by an iPEPS with a 
2-sublattice unit cell consisting of two local tensors (four sites). 

We compute physical expectation values using a variant~\cite{tnref-corboz14} 
of the corner-transfer-matrix (CTM) algorithm~\cite{tnref-nishino96,
tnref-orus09}. The corner matrices have their own boundary bond dimension, 
$\chi$, which should be taken to be sufficiently large ($\chi(D) > D^2$) that 
the error due to the use of finite $\chi$ is negligible compared to the error 
due to the finite value of $D$. To increase the efficiency of our calculations 
we exploit the global U(1) symmetry of the model \cite{tnref-singh11,
tnref-bauer11}.

Given an initial iPEPS, we obtain an approximate ground state either by 
projecting the starting state using imaginary-time evolution or by direct 
minimization of the energy using the variational-update method of 
Ref.~\cite{tnref-corboz16b}. In the former approach, the projection operator 
is decomposed into a series of two-body gates. Application of a single gate 
increases the dimension of the bond connecting the two tensors in question, 
which then has to be truncated back to $D$. This process may be performed 
using the simple-update method~\cite{tnref-jiang08}, in which the truncation 
of a bond index is based on a local approximation of the state, or by the 
more accurate but computationally more expensive full-update algorithm 
\cite{tnref-jordan08,tnref-corboz10,tnref-phien15}, where the entire 
many-body state is taken into account for the truncation.

\begin{figure}[t!]
\centering
\includegraphics[width=\columnwidth]{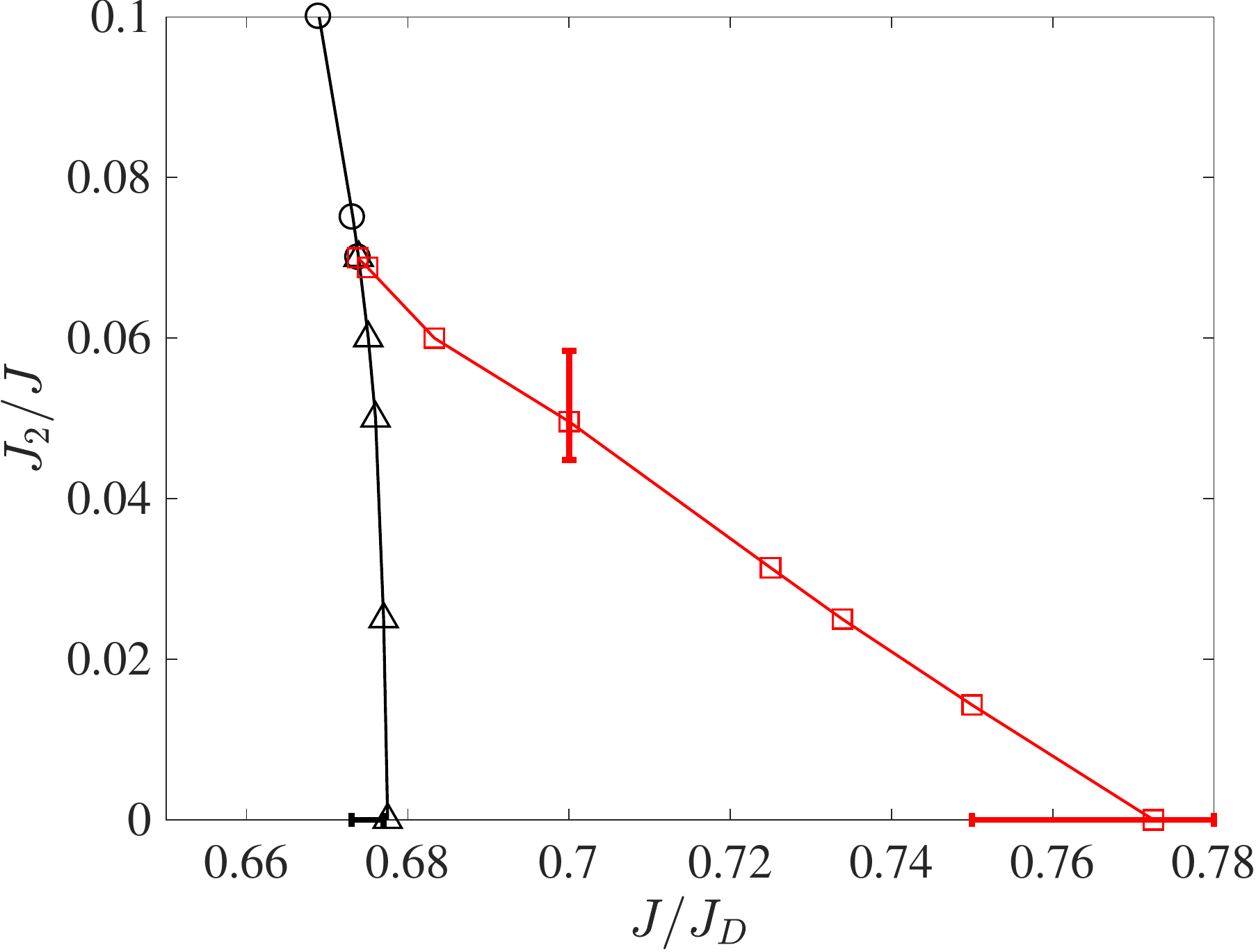}
\caption{Phase boundaries obtained by simple-update iPEPS calculations. The 
error bars illustrated for three data points are representative of all others. 
These were obtained from extrapolated ($D \rightarrow \infty$) full-update 
calculations at $J_2/J = 0$ and from variational-update calculations at $J/J_D
 = 0.7$. Results for $J_2/J = 0$ were taken from Ref.~\cite{CorbozMila13}.}
\label{fig:phase_diagram_peps_zoom}
\end{figure}

To construct the phase diagram shown in Fig.~\ref{fig:phase_diagram_peps}, we 
employed simple-update calculations at a fixed bond dimension $D = 10$, which 
already provide a good estimate of the phase boundaries in the limit of 
infinite $D$, as we show below. We computed the transition points along 
several horizontal and vertical cuts through the phase diagram. Working at a 
fixed value of $J_2/J$ for a horizontal cut, the critical coupling $J_c/J_D$ 
was determined by locating the intersection point where the energies of states 
initialized in the two adjacent phases intersect (making use of the hysteresis 
in the vicinity of a first-order phase transition). We note that, because the 
ground-state energy in the dimer singlet-product phase is known exactly, the 
fixed-$D$ estimate of the phase boundary between this dimer phase and either 
of the other phases (antiferromagnetic or plaquette) shifts to smaller values 
of $J/J_D$ with increasing $D$.

To determine the accuracy of the fixed-$D$ simple-update phase diagram, we have 
executed additional variational-update calculations followed by extrapolations 
to the $D \rightarrow \infty$ limit, where our results should be exact, along 
several cuts in the parameter space. Extrapolations in this case were performed 
on the basis of the truncation error \cite{tnref-corboz16}. By comparison with 
the $D = 10$ simple-update phase boundaries along four horizontal cuts, taken 
at $J_2/J = 0.25$, $0.50$, $0.75$, and $1.00$, we observe that the phase 
boundary for the transition from the singlet-product to the antiferromagnetic 
state, displayed in Fig.~\ref{fig:phase_diagram_peps}, agrees with the 
variational-update $D \rightarrow \infty$ phase boundary up to the first 
four digits. The uncertainty in the phase boundaries of the plaquette phase 
is somewhat larger, and is represented by the error bars on three of the 
points shown in Fig.~\ref{fig:phase_diagram_peps_zoom}, which were obtained 
from detailed studies along two horizontal cuts at $J_2/J = 0$ and a vertical 
cut at $J/J_D = 0.7$. We comment that the error bars for the transition from 
the plaquette to the antiferromagnetic phase are the largest because this 
transition appears to be only weakly first-order. The thickness of the lines 
marking the phase boundaries in Fig.~\ref{fig:phase_diagram_peps} was 
determined on the basis of the error bars shown in 
Fig.~\ref{fig:phase_diagram_peps_zoom}.

\section{Interpolation of High-Temperature Series Expansions}
\label{sec:appHighTser}

As its name implies, the aim of a HTSE is to express the magnetic 
susceptibility and specific heat in powers of the inverse temperature, 
\begin{equation}
\chi(T) = \sum_{n=0}^{M} \chi_n \,T^{-n}, \qquad
C(T) = \sum_{n=0}^{M} C_n \,T^{-n},
\label{eq:PlainSeries}
\end{equation}
in order to obtain results exact in the high-$T$ limit and systematic 
approximations elsewhere. We began our study by using the methods and 
code described in Ref.~\cite{res:LSR14} to generate series to order 
$M = 10$ for $\chi(T)$ and $C(T)$ in the Shastry-Sutherland model. 
However, the truncated bare series of Eq.~(\ref{eq:PlainSeries}) diverge in 
the low-temperature regime, which is the focus of the present study. The 
conventional solution to this divergence is the use of Pad\'e approximants 
(reviewed in Ref.~\cite{Guttmann1989}), but this approach is completely 
unsuitable here because it always yields a power-law low-temperature 
behavior, rather than the exponentially activated behavior characteristic 
of a gapped model (Sec.~\ref{Physical results}). 

Thus we adopt a simple approach to constructing an interpolation scheme, 
which is to exchange variables in order to work with an expansion in terms 
of exponential functions, ${\rm e}^{-\Delta/T}$, containing a gap parameter 
$\Delta$. We comment that several similar but more sophisticated schemes 
have been proposed recently \cite{PhysRevB.63.134409,PhysRevLett.114.057201,
PhysRevE.95.042110}. Here we take the additional step of incorporating the 
known leading high-temperature asymptotics into the ansatz to obtain 
\begin{eqnarray}
\chi_T(T) & = & \frac{1}{T} \, \sum_{n=1}^{M_\chi} \tilde{\chi}_n \, 
{\rm e}^{-n\,\Delta/T}, \label{eq:expTSeries} \\
C_T(T) & = & \frac{1}{T^2} \, \sum_{n=1}^{M_C} \tilde{C}_n \,{\rm e}^{-n\,\Delta/T}.
\label{eq:expTSeriesC}
\end{eqnarray}
Because the exponential functions within the sum decay faster than any power 
law, the expressions (\ref{eq:expTSeries}) and (\ref{eq:expTSeriesC}) have 
well-defined low-temperature behavior and tend to zero as $T \to 0$. The 
coefficients $\tilde{\chi}_n$ and $\tilde{C}_n$ may be determined by 
demanding that the Taylor expansions of Eqs.~(\ref{eq:expTSeries}) and 
(\ref{eq:expTSeriesC}) match the corresponding coefficients in 
Eq.~(\ref{eq:PlainSeries}). 

\begin{figure}[t]
\centering
\includegraphics[width=0.48\columnwidth]{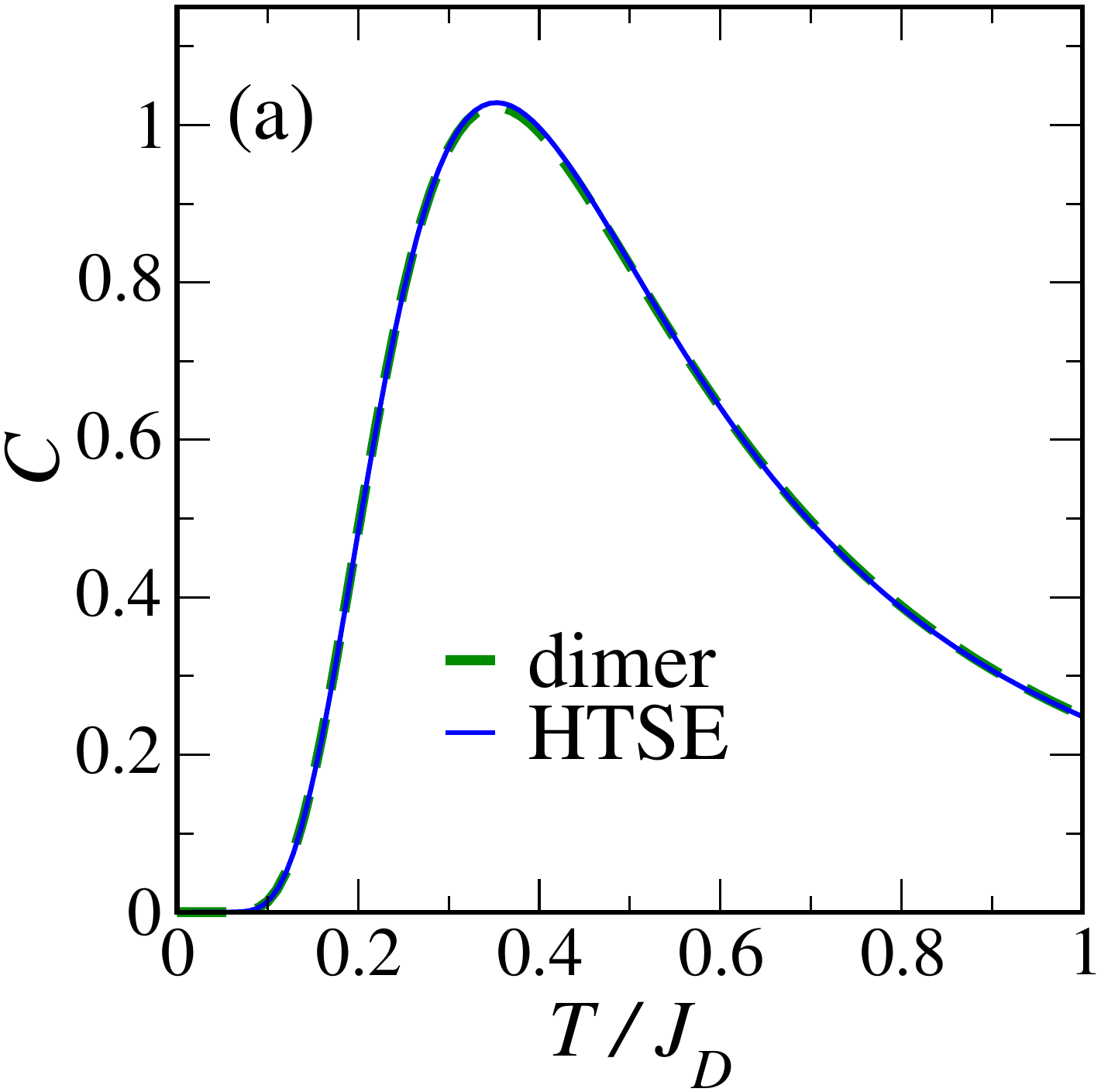}
\includegraphics[width=0.49\columnwidth]{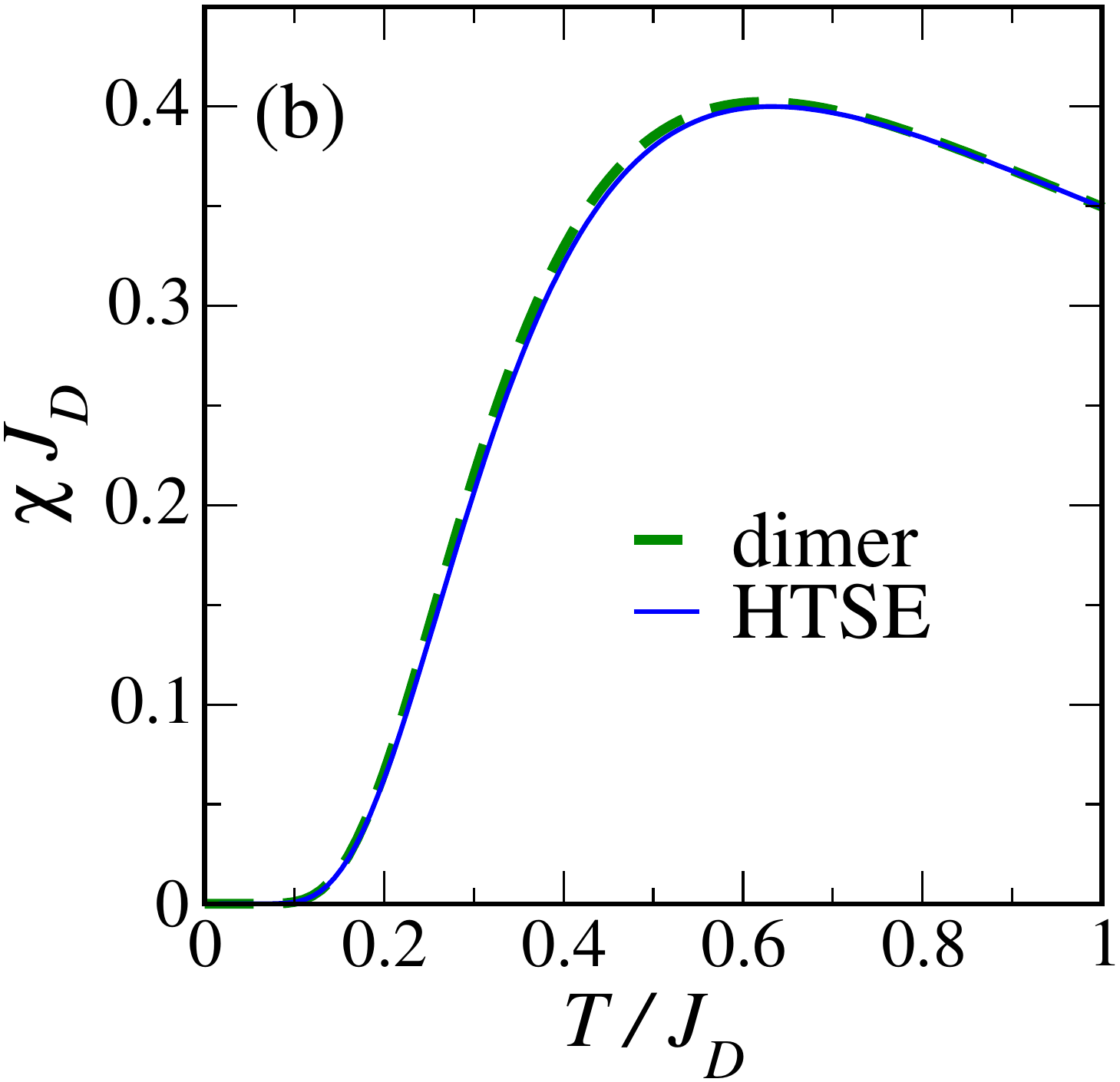}
\caption{(a) Magnetic specific heat, $C(T)$, and (b) susceptibility, $\chi(T)$, 
of an isolated spin-1/2 dimer, shown for comparison with the interpolated 
tenth-order HTSE result.}
\label{fig:CchiDimer}
\end{figure}

For the susceptibility, indeed we determine the $M_\chi = M$ 
coefficients $\tilde{\chi}_n$ in this manner. For the specific heat, we 
can obtain additional constraints, following Refs.~\cite{PhysRevB.63.134409,
PhysRevLett.114.057201,PhysRevE.95.042110}, by further imposing two sum rules, 
one for the ground-state energy,
\begin{equation}
E_0 = - \int\limits_0^\infty {\rm d}T \, C(T),
\label{eq:CsumRule}
\end{equation}
and one for the total entropy per dimer of a spin-1/2 system,
\begin{equation}
\int\limits_0^\infty {\rm d}T \, \frac{C(T)}{T} = 2\, \ln 2.
\label{eq:SsumRule}
\end{equation}
Performing the relevant integrals in Eq.~(\ref{eq:expTSeriesC}) yields the 
two additional linear equations
\begin{equation}
\sum_{n=1}^{M_C}  \frac{\tilde{C}_n}{n\,\Delta} = -E_0 \quad {\rm and} \quad
\sum_{n=1}^{M_C}  \frac{\tilde{C}_n}{(n\,\Delta)^2} =  2\,\ln2
\label{eq:expCTcons}
\end{equation}
constraining the $M_C = M + 1$ coefficients $\tilde{C}_n$. We comment that 
the respective relations between the coefficients $\chi_n$, $C_n$ and
$\tilde{\chi}_n$, $\tilde{C}_n$ are highly non-trivial. In particular, the 
individual coefficients $\tilde{\chi}_n$ and $\tilde{C}_n$ are not constrained 
to converge when $M \to \infty$, i.e.~the entire procedure should be
considered only as an interpolation between the low- and high-temperature 
limits using a finite-order approximation to the latter.

This interpolation procedure makes use of two additional parameters. 
One is the ground-state energy per dimer unit cell, which as noted in 
Sec.~\ref{The models} is known exactly for the Shastry-Sutherland model 
in its singlet-product state \cite{ShaSu81}, namely $E_0 = - {\textstyle 
\frac{3}{4}} \,J_D$. The other is a value for the gap at any given coupling 
ratio $J/J_D$, and in our present HTSE studies we have used the values of 
the gap obtained by ED for the $N = 36$ cluster, which are shown in 
Fig.~\ref{fig:Gap36}. We comment again that for $J/J_D \le 0.6$ the lowest 
excitation is indeed the one-particle triplet mode, and thus that no 
distinction is required between the gaps used for the susceptibility and 
specific-heat expansions.

We illustrate the efficacy of the HTSE interpolation procedure by using 
the example of the isolated dimer, i.e.~the case $J = 0 = J_2$ in 
Eq.~(\ref{extended}). Exact results for $\chi(T)$ and $C(T)$ of a 
single dimer are known analytically \cite{Bleaney451,PhysRevB.61.9558,
PhysRevB.74.174421,rus1}, and in fact one may observe explicitly that 
low-temperature expansions of the exact expressions correspond precisely 
to the ansatz used in Eqs.~(\ref{eq:expTSeries}) and (\ref{eq:expTSeriesC}). 
Figure~\ref{fig:CchiDimer} compares these exact results for $\chi(T)$ and 
$C(T)$, shown already in Fig.~\ref{fig:cchi}, with the interpolated 
tenth-order HTSE. Although the overall level of agreement could be 
classified as excellent, some deviations can be observed upon close 
inspection. We remark that, even in the isolated-dimer limit, the energy 
and entropy sum rules (\ref{eq:expCTcons}) are essential to stabilize the 
interpolation of $C(T)$ at lower temperatures, most notably around its 
maximum. In $\chi(T)$, which is less well constrained, minor deviations 
are evident in the temperature scale [cf.~Figs.~\ref{fig:Cchi}(b), 
\ref{fig:Cchi}(d), and \ref{fig:Cchi}(f)] as well as in the magnitude. 

\bibliography{ShaSuQMC,tnrefs}

\begin{thebibliography}{72}%
\makeatletter
\providecommand \@ifxundefined [1]{%
 \@ifx{#1\undefined}
}%
\providecommand \@ifnum [1]{%
 \ifnum #1\expandafter \@firstoftwo
 \else \expandafter \@secondoftwo
 \fi
}%
\providecommand \@ifx [1]{%
 \ifx #1\expandafter \@firstoftwo
 \else \expandafter \@secondoftwo
 \fi
}%
\providecommand \natexlab [1]{#1}%
\providecommand \enquote  [1]{``#1''}%
\providecommand \bibnamefont  [1]{#1}%
\providecommand \bibfnamefont [1]{#1}%
\providecommand \citenamefont [1]{#1}%
\providecommand \href@noop [0]{\@secondoftwo}%
\providecommand \href [0]{\begingroup \@sanitize@url \@href}%
\providecommand \@href[1]{\@@startlink{#1}\@@href}%
\providecommand \@@href[1]{\endgroup#1\@@endlink}%
\providecommand \@sanitize@url [0]{\catcode `\\12\catcode `\$12\catcode
  `\&12\catcode `\#12\catcode `\^12\catcode `\_12\catcode `\%12\relax}%
\providecommand \@@startlink[1]{}%
\providecommand \@@endlink[0]{}%
\providecommand \url  [0]{\begingroup\@sanitize@url \@url }%
\providecommand \@url [1]{\endgroup\@href {#1}{\urlprefix }}%
\providecommand \urlprefix  [0]{URL }%
\providecommand \Eprint [0]{\href }%
\providecommand \doibase [0]{http://dx.doi.org/}%
\providecommand \selectlanguage [0]{\@gobble}%
\providecommand \bibinfo  [0]{\@secondoftwo}%
\providecommand \bibfield  [0]{\@secondoftwo}%
\providecommand \translation [1]{[#1]}%
\providecommand \BibitemOpen [0]{}%
\providecommand \bibitemStop [0]{}%
\providecommand \bibitemNoStop [0]{.\EOS\space}%
\providecommand \EOS [0]{\spacefactor3000\relax}%
\providecommand \BibitemShut  [1]{\csname bibitem#1\endcsname}%
\let\auto@bib@innerbib\@empty
\bibitem [{\citenamefont {Richter}\ \emph {et~al.}(2004)\citenamefont
  {Richter}, \citenamefont {Schulenburg},\ and\ \citenamefont
  {Honecker}}]{Richter2004}%
  \BibitemOpen
  \bibfield  {author} {\bibinfo {author} {\bibfnamefont {Johannes}\
  \bibnamefont {Richter}}, \bibinfo {author} {\bibfnamefont {J\"org}\
  \bibnamefont {Schulenburg}}, \ and\ \bibinfo {author} {\bibfnamefont
  {Andreas}\ \bibnamefont {Honecker}},\ }\enquote {\bibinfo {title} {Quantum
  magnetism in two dimensions: From semi-classical {N{\'e}el} order to magnetic
  disorder},}\ in\ \href {\doibase 10.1007/BFb0119592} {\emph {\bibinfo
  {booktitle} {Quantum Magnetism}}},\ \bibinfo {editor} {edited by\ \bibinfo
  {editor} {\bibfnamefont {Ulrich}\ \bibnamefont {Schollw\"ock}}, \bibinfo
  {editor} {\bibfnamefont {Johannes}\ \bibnamefont {Richter}}, \bibinfo
  {editor} {\bibfnamefont {Damian J.~J.}\ \bibnamefont {Farnell}}, \ and\
  \bibinfo {editor} {\bibfnamefont {Raymod~F.}\ \bibnamefont {Bishop}}}\
  (\bibinfo  {publisher} {Springer Berlin Heidelberg},\ \bibinfo {address}
  {Berlin, Heidelberg},\ \bibinfo {year} {2004})\ pp.\ \bibinfo {pages}
  {85--153}\BibitemShut {NoStop}%
\bibitem [{\citenamefont {Balents}(2010)}]{Balents10}%
  \BibitemOpen
  \bibfield  {author} {\bibinfo {author} {\bibfnamefont {Leon}\ \bibnamefont
  {Balents}},\ }\bibfield  {title} {\enquote {\bibinfo {title} {Spin liquids in
  frustrated magnets},}\ }\href {\doibase 10.1038/nature08917} {\bibfield
  {journal} {\bibinfo  {journal} {Nature}\ }\textbf {\bibinfo {volume} {464}},\
  \bibinfo {pages} {199--208} (\bibinfo {year} {2010})}\BibitemShut {NoStop}%
\bibitem [{\citenamefont {Lacroix}\ \emph {et~al.}(2011)\citenamefont
  {Lacroix}, \citenamefont {Mendels},\ and\ \citenamefont {Mila}}]{HFMbook}%
  \BibitemOpen
  \bibfield  {author} {\bibinfo {author} {\bibfnamefont {Claudine}\
  \bibnamefont {Lacroix}}, \bibinfo {author} {\bibfnamefont {Philippe}\
  \bibnamefont {Mendels}}, \ and\ \bibinfo {author} {\bibfnamefont
  {Fr{\'e}d{\'e}ric}\ \bibnamefont {Mila}},\ }\href@noop {} {\emph {\bibinfo
  {title} {Introduction to Frustrated Magnetism: Materials, Experiments,
  Theory}}},\ \bibinfo {series} {Springer Series in Solid-State Sciences},
  Vol.\ \bibinfo {volume} {164}\ (\bibinfo  {publisher} {Springer Berlin
  Heidelberg},\ \bibinfo {address} {Berlin, Heidelberg},\ \bibinfo {year}
  {2011})\BibitemShut {NoStop}%
\bibitem [{\citenamefont {Diep}(2013)}]{DIEPbook}%
  \BibitemOpen
  \bibfield  {author} {\bibinfo {author} {\bibfnamefont {H.~T.}\ \bibnamefont
  {Diep}},\ }\href@noop {} {\emph {\bibinfo {title} {Frustrated Spin Systems,
  2nd edition}}}\ (\bibinfo  {publisher} {World Scientific},\ \bibinfo
  {address} {Singapore},\ \bibinfo {year} {2013})\BibitemShut {NoStop}%
\bibitem [{\citenamefont {Kim}\ and\ \citenamefont
  {Troyer}(1998)}]{PhysRevLett.80.2705}%
  \BibitemOpen
  \bibfield  {author} {\bibinfo {author} {\bibfnamefont {Jae-Kwon}\
  \bibnamefont {Kim}}\ and\ \bibinfo {author} {\bibfnamefont {Matthias}\
  \bibnamefont {Troyer}},\ }\bibfield  {title} {\enquote {\bibinfo {title} {Low
  temperature behavior and crossovers of the square lattice quantum
  {Heisenberg} antiferromagnet},}\ }\href {\doibase
  10.1103/PhysRevLett.80.2705} {\bibfield  {journal} {\bibinfo  {journal}
  {Phys. Rev. Lett.}\ }\textbf {\bibinfo {volume} {80}},\ \bibinfo {pages}
  {2705--2708} (\bibinfo {year} {1998})}\BibitemShut {NoStop}%
\bibitem [{\citenamefont {Harada}\ \emph {et~al.}(1998)\citenamefont {Harada},
  \citenamefont {Troyer},\ and\ \citenamefont {Kawashima}}]{HTN98}%
  \BibitemOpen
  \bibfield  {author} {\bibinfo {author} {\bibfnamefont {Kenji}\ \bibnamefont
  {Harada}}, \bibinfo {author} {\bibfnamefont {Matthias}\ \bibnamefont
  {Troyer}}, \ and\ \bibinfo {author} {\bibfnamefont {Naoki}\ \bibnamefont
  {Kawashima}},\ }\bibfield  {title} {\enquote {\bibinfo {title} {The
  two-dimensional {$S=1$} quantum {Heisenberg} antiferromagnet at finite
  temperatures},}\ }\href {\doibase 10.1143/JPSJ.67.1130} {\bibfield  {journal}
  {\bibinfo  {journal} {J. Phys. Soc. Jpn.}\ }\textbf {\bibinfo {volume}
  {67}},\ \bibinfo {pages} {1130--1133} (\bibinfo {year} {1998})}\BibitemShut
  {NoStop}%
\bibitem [{\citenamefont {Shastry}\ and\ \citenamefont
  {Sutherland}(1981)}]{ShaSu81}%
  \BibitemOpen
  \bibfield  {author} {\bibinfo {author} {\bibfnamefont {B.~Sriram}\
  \bibnamefont {Shastry}}\ and\ \bibinfo {author} {\bibfnamefont {Bill}\
  \bibnamefont {Sutherland}},\ }\bibfield  {title} {\enquote {\bibinfo {title}
  {Exact ground state of a quantum mechanical antiferromagnet},}\ }\href
  {\doibase 0378-4363(81)90838-X} {\bibfield  {journal} {\bibinfo  {journal}
  {Physica B+C}\ }\textbf {\bibinfo {volume} {108}},\ \bibinfo {pages}
  {1069--1070} (\bibinfo {year} {1981})}\BibitemShut {NoStop}%
\bibitem [{\citenamefont {Liao}\ \emph {et~al.}(2017)\citenamefont {Liao},
  \citenamefont {Xie}, \citenamefont {Chen}, \citenamefont {Liu}, \citenamefont
  {Xie}, \citenamefont {Huang}, \citenamefont {Normand},\ and\ \citenamefont
  {Xiang}}]{Liao17}%
  \BibitemOpen
  \bibfield  {author} {\bibinfo {author} {\bibfnamefont {H.~J.}\ \bibnamefont
  {Liao}}, \bibinfo {author} {\bibfnamefont {Z.~Y.}\ \bibnamefont {Xie}},
  \bibinfo {author} {\bibfnamefont {J.}~\bibnamefont {Chen}}, \bibinfo {author}
  {\bibfnamefont {Z.~Y.}\ \bibnamefont {Liu}}, \bibinfo {author} {\bibfnamefont
  {H.~D.}\ \bibnamefont {Xie}}, \bibinfo {author} {\bibfnamefont {R.~Z.}\
  \bibnamefont {Huang}}, \bibinfo {author} {\bibfnamefont {B.}~\bibnamefont
  {Normand}}, \ and\ \bibinfo {author} {\bibfnamefont {T.}~\bibnamefont
  {Xiang}},\ }\bibfield  {title} {\enquote {\bibinfo {title} {Gapless
  spin-liquid ground state in the {$S = 1/2$} kagome antiferromagnet},}\ }\href
  {\doibase 10.1103/PhysRevLett.118.137202} {\bibfield  {journal} {\bibinfo
  {journal} {Phys. Rev. Lett.}\ }\textbf {\bibinfo {volume} {118}},\ \bibinfo
  {pages} {137202} (\bibinfo {year} {2017})}\BibitemShut {NoStop}%
\bibitem [{\citenamefont {Kageyama}\ \emph {et~al.}(1999)\citenamefont
  {Kageyama}, \citenamefont {Yoshimura}, \citenamefont {Stern}, \citenamefont
  {Mushnikov}, \citenamefont {Onizuka}, \citenamefont {Kato}, \citenamefont
  {Kosuge}, \citenamefont {Slichter}, \citenamefont {Goto},\ and\ \citenamefont
  {Ueda}}]{PhysRevLett.82.3168}%
  \BibitemOpen
  \bibfield  {author} {\bibinfo {author} {\bibfnamefont {H.}~\bibnamefont
  {Kageyama}}, \bibinfo {author} {\bibfnamefont {K.}~\bibnamefont {Yoshimura}},
  \bibinfo {author} {\bibfnamefont {R.}~\bibnamefont {Stern}}, \bibinfo
  {author} {\bibfnamefont {N.~V.}\ \bibnamefont {Mushnikov}}, \bibinfo {author}
  {\bibfnamefont {K.}~\bibnamefont {Onizuka}}, \bibinfo {author} {\bibfnamefont
  {M.}~\bibnamefont {Kato}}, \bibinfo {author} {\bibfnamefont {K.}~\bibnamefont
  {Kosuge}}, \bibinfo {author} {\bibfnamefont {C.~P.}\ \bibnamefont
  {Slichter}}, \bibinfo {author} {\bibfnamefont {T.}~\bibnamefont {Goto}}, \
  and\ \bibinfo {author} {\bibfnamefont {Y.}~\bibnamefont {Ueda}},\ }\bibfield
  {title} {\enquote {\bibinfo {title} {Exact dimer ground state and quantized
  magnetization plateaus in the two-dimensional spin system
  {SrCu$_{2}$(BO$_{3}$)$_{2}$}},}\ }\href {\doibase
  10.1103/PhysRevLett.82.3168} {\bibfield  {journal} {\bibinfo  {journal}
  {Phys. Rev. Lett.}\ }\textbf {\bibinfo {volume} {82}},\ \bibinfo {pages}
  {3168--3171} (\bibinfo {year} {1999})}\BibitemShut {NoStop}%
\bibitem [{\citenamefont {Onizuka}\ \emph {et~al.}(2000)\citenamefont
  {Onizuka}, \citenamefont {Kageyama}, \citenamefont {Narumi}, \citenamefont
  {Kindo}, \citenamefont {Ueda},\ and\ \citenamefont {Goto}}]{Onizuka00}%
  \BibitemOpen
  \bibfield  {author} {\bibinfo {author} {\bibfnamefont {K.}~\bibnamefont
  {Onizuka}}, \bibinfo {author} {\bibfnamefont {H.}~\bibnamefont {Kageyama}},
  \bibinfo {author} {\bibfnamefont {Y.}~\bibnamefont {Narumi}}, \bibinfo
  {author} {\bibfnamefont {K.}~\bibnamefont {Kindo}}, \bibinfo {author}
  {\bibfnamefont {Y.}~\bibnamefont {Ueda}}, \ and\ \bibinfo {author}
  {\bibfnamefont {T.}~\bibnamefont {Goto}},\ }\bibfield  {title} {\enquote
  {\bibinfo {title} {1/3 magnetization plateau in {SrCu$_{2}$(BO$_{3}$)$_{2}$}
  - stripe order of excited triplets -},}\ }\href {\doibase
  10.1143/JPSJ.69.1016} {\bibfield  {journal} {\bibinfo  {journal} {J. Phys.
  Soc. Jpn.}\ }\textbf {\bibinfo {volume} {69}},\ \bibinfo {pages} {1016--1018}
  (\bibinfo {year} {2000})}\BibitemShut {NoStop}%
\bibitem [{\citenamefont {Kodama}\ \emph {et~al.}(2002)\citenamefont {Kodama},
  \citenamefont {Takigawa}, \citenamefont {Horvati{\'c}}, \citenamefont
  {Berthier}, \citenamefont {Kageyama}, \citenamefont {Ueda}, \citenamefont
  {Miyahara}, \citenamefont {Becca},\ and\ \citenamefont {Mila}}]{Kodama02}%
  \BibitemOpen
  \bibfield  {author} {\bibinfo {author} {\bibfnamefont {K.}~\bibnamefont
  {Kodama}}, \bibinfo {author} {\bibfnamefont {M.}~\bibnamefont {Takigawa}},
  \bibinfo {author} {\bibfnamefont {M.}~\bibnamefont {Horvati{\'c}}}, \bibinfo
  {author} {\bibfnamefont {C.}~\bibnamefont {Berthier}}, \bibinfo {author}
  {\bibfnamefont {H.}~\bibnamefont {Kageyama}}, \bibinfo {author}
  {\bibfnamefont {Y.}~\bibnamefont {Ueda}}, \bibinfo {author} {\bibfnamefont
  {S.}~\bibnamefont {Miyahara}}, \bibinfo {author} {\bibfnamefont
  {F.}~\bibnamefont {Becca}}, \ and\ \bibinfo {author} {\bibfnamefont
  {F.}~\bibnamefont {Mila}},\ }\bibfield  {title} {\enquote {\bibinfo {title}
  {Magnetic superstructure in the two-dimensional quantum antiferromagnet
  {SrCu$_2$(BO$_3$)$_2$}},}\ }\href {\doibase 10.1126/science.1075045}
  {\bibfield  {journal} {\bibinfo  {journal} {Science}\ }\textbf {\bibinfo
  {volume} {298}},\ \bibinfo {pages} {395--399} (\bibinfo {year}
  {2002})}\BibitemShut {NoStop}%
\bibitem [{\citenamefont {Sebastian}\ \emph {et~al.}(2008)\citenamefont
  {Sebastian}, \citenamefont {Harrison}, \citenamefont {Sengupta},
  \citenamefont {Batista}, \citenamefont {Francoual}, \citenamefont {Palm},
  \citenamefont {Murphy}, \citenamefont {Marcano}, \citenamefont {Dabkowska},\
  and\ \citenamefont {Gaulin}}]{Sebastian08}%
  \BibitemOpen
  \bibfield  {author} {\bibinfo {author} {\bibfnamefont {Suchitra~E.}\
  \bibnamefont {Sebastian}}, \bibinfo {author} {\bibfnamefont {N.}~\bibnamefont
  {Harrison}}, \bibinfo {author} {\bibfnamefont {P.}~\bibnamefont {Sengupta}},
  \bibinfo {author} {\bibfnamefont {C.~D.}\ \bibnamefont {Batista}}, \bibinfo
  {author} {\bibfnamefont {S.}~\bibnamefont {Francoual}}, \bibinfo {author}
  {\bibfnamefont {E.}~\bibnamefont {Palm}}, \bibinfo {author} {\bibfnamefont
  {T.}~\bibnamefont {Murphy}}, \bibinfo {author} {\bibfnamefont
  {N.}~\bibnamefont {Marcano}}, \bibinfo {author} {\bibfnamefont {H.~A.}\
  \bibnamefont {Dabkowska}}, \ and\ \bibinfo {author} {\bibfnamefont {B.~D.}\
  \bibnamefont {Gaulin}},\ }\bibfield  {title} {\enquote {\bibinfo {title}
  {Fractalization drives crystalline states in a frustrated spin system},}\
  }\href {\doibase 10.1073/pnas.0804320105} {\bibfield  {journal} {\bibinfo
  {journal} {Proceedings of the National Academy of Sciences}\ }\textbf
  {\bibinfo {volume} {105}},\ \bibinfo {pages} {20157--20160} (\bibinfo {year}
  {2008})}\BibitemShut {NoStop}%
\bibitem [{\citenamefont {Takigawa}\ and\ \citenamefont
  {Mila}(2011)}]{Takigawa2011}%
  \BibitemOpen
  \bibfield  {author} {\bibinfo {author} {\bibfnamefont {Masashi}\ \bibnamefont
  {Takigawa}}\ and\ \bibinfo {author} {\bibfnamefont {Fr{\'e}d{\'e}ric}\
  \bibnamefont {Mila}},\ }\enquote {\bibinfo {title} {Magnetization
  plateaus},}\ in\ \href {\doibase 10.1007/978-3-642-10589-0_10} {\emph
  {\bibinfo {booktitle} {Introduction to Frustrated Magnetism: Materials,
  Experiments, Theory}}},\ \bibinfo {editor} {edited by\ \bibinfo {editor}
  {\bibfnamefont {Claudine}\ \bibnamefont {Lacroix}}, \bibinfo {editor}
  {\bibfnamefont {Philippe}\ \bibnamefont {Mendels}}, \ and\ \bibinfo {editor}
  {\bibfnamefont {Fr{\'e}d{\'e}ric}\ \bibnamefont {Mila}}}\ (\bibinfo
  {publisher} {Springer Berlin Heidelberg},\ \bibinfo {address} {Berlin,
  Heidelberg},\ \bibinfo {year} {2011})\ pp.\ \bibinfo {pages}
  {241--267}\BibitemShut {NoStop}%
\bibitem [{\citenamefont {Takigawa}\ \emph {et~al.}(2013)\citenamefont
  {Takigawa}, \citenamefont {Horvati\ifmmode~\acute{c}\else \'{c}\fi{}},
  \citenamefont {Waki}, \citenamefont {Kr\"amer}, \citenamefont {Berthier},
  \citenamefont {L\'evy-Bertrand}, \citenamefont {Sheikin}, \citenamefont
  {Kageyama}, \citenamefont {Ueda},\ and\ \citenamefont {Mila}}]{Takigawa12}%
  \BibitemOpen
  \bibfield  {author} {\bibinfo {author} {\bibfnamefont {M.}~\bibnamefont
  {Takigawa}}, \bibinfo {author} {\bibfnamefont {M.}~\bibnamefont
  {Horvati\ifmmode~\acute{c}\else \'{c}\fi{}}}, \bibinfo {author}
  {\bibfnamefont {T.}~\bibnamefont {Waki}}, \bibinfo {author} {\bibfnamefont
  {S.}~\bibnamefont {Kr\"amer}}, \bibinfo {author} {\bibfnamefont
  {C.}~\bibnamefont {Berthier}}, \bibinfo {author} {\bibfnamefont
  {F.}~\bibnamefont {L\'evy-Bertrand}}, \bibinfo {author} {\bibfnamefont
  {I.}~\bibnamefont {Sheikin}}, \bibinfo {author} {\bibfnamefont
  {H.}~\bibnamefont {Kageyama}}, \bibinfo {author} {\bibfnamefont
  {Y.}~\bibnamefont {Ueda}}, \ and\ \bibinfo {author} {\bibfnamefont
  {F.}~\bibnamefont {Mila}},\ }\bibfield  {title} {\enquote {\bibinfo {title}
  {Incomplete devil's staircase in the magnetization curve of
  {SrCu$_{2}$(BO$_{3}$)$_{2}$}},}\ }\href {\doibase
  10.1103/PhysRevLett.110.067210} {\bibfield  {journal} {\bibinfo  {journal}
  {Phys. Rev. Lett.}\ }\textbf {\bibinfo {volume} {110}},\ \bibinfo {pages}
  {067210} (\bibinfo {year} {2013})}\BibitemShut {NoStop}%
\bibitem [{\citenamefont {Jaime}\ \emph {et~al.}(2012)\citenamefont {Jaime},
  \citenamefont {Daou}, \citenamefont {Crooker}, \citenamefont {Weickert},
  \citenamefont {Uchida}, \citenamefont {Feiguin}, \citenamefont {Batista},
  \citenamefont {Dabkowska},\ and\ \citenamefont {Gaulin}}]{Jaime12}%
  \BibitemOpen
  \bibfield  {author} {\bibinfo {author} {\bibfnamefont {Marcelo}\ \bibnamefont
  {Jaime}}, \bibinfo {author} {\bibfnamefont {Ramzy}\ \bibnamefont {Daou}},
  \bibinfo {author} {\bibfnamefont {Scott~A.}\ \bibnamefont {Crooker}},
  \bibinfo {author} {\bibfnamefont {Franziska}\ \bibnamefont {Weickert}},
  \bibinfo {author} {\bibfnamefont {Atsuko}\ \bibnamefont {Uchida}}, \bibinfo
  {author} {\bibfnamefont {Adrian~E.}\ \bibnamefont {Feiguin}}, \bibinfo
  {author} {\bibfnamefont {Cristian~D.}\ \bibnamefont {Batista}}, \bibinfo
  {author} {\bibfnamefont {Hanna~A.}\ \bibnamefont {Dabkowska}}, \ and\
  \bibinfo {author} {\bibfnamefont {Bruce~D.}\ \bibnamefont {Gaulin}},\
  }\bibfield  {title} {\enquote {\bibinfo {title} {Magnetostriction and
  magnetic texture to 100.75 {T}esla in frustrated {SrCu$_2$(BO$_3$)$_2$}},}\
  }\href {\doibase 10.1073/pnas.1200743109} {\bibfield  {journal} {\bibinfo
  {journal} {Proceedings of the National Academy of Sciences}\ }\textbf
  {\bibinfo {volume} {109}},\ \bibinfo {pages} {12404--12407} (\bibinfo {year}
  {2012})}\BibitemShut {NoStop}%
\bibitem [{\citenamefont {Matsuda}\ \emph {et~al.}(2013)\citenamefont
  {Matsuda}, \citenamefont {Abe}, \citenamefont {Takeyama}, \citenamefont
  {Kageyama}, \citenamefont {Corboz}, \citenamefont {Honecker}, \citenamefont
  {Manmana}, \citenamefont {Foltin}, \citenamefont {Schmidt},\ and\
  \citenamefont {Mila}}]{PhysRevLett.111.137204}%
  \BibitemOpen
  \bibfield  {author} {\bibinfo {author} {\bibfnamefont {Y.~H.}\ \bibnamefont
  {Matsuda}}, \bibinfo {author} {\bibfnamefont {N.}~\bibnamefont {Abe}},
  \bibinfo {author} {\bibfnamefont {S.}~\bibnamefont {Takeyama}}, \bibinfo
  {author} {\bibfnamefont {H.}~\bibnamefont {Kageyama}}, \bibinfo {author}
  {\bibfnamefont {P.}~\bibnamefont {Corboz}}, \bibinfo {author} {\bibfnamefont
  {A.}~\bibnamefont {Honecker}}, \bibinfo {author} {\bibfnamefont {S.~R.}\
  \bibnamefont {Manmana}}, \bibinfo {author} {\bibfnamefont {G.~R.}\
  \bibnamefont {Foltin}}, \bibinfo {author} {\bibfnamefont {K.~P.}\
  \bibnamefont {Schmidt}}, \ and\ \bibinfo {author} {\bibfnamefont
  {F.}~\bibnamefont {Mila}},\ }\bibfield  {title} {\enquote {\bibinfo {title}
  {Magnetization of {SrCu$_{2}$(BO$_{3}$)$_{2}$} in ultrahigh magnetic fields
  up to 118 {T}},}\ }\href {\doibase 10.1103/PhysRevLett.111.137204} {\bibfield
   {journal} {\bibinfo  {journal} {Phys. Rev. Lett.}\ }\textbf {\bibinfo
  {volume} {111}},\ \bibinfo {pages} {137204} (\bibinfo {year}
  {2013})}\BibitemShut {NoStop}%
\bibitem [{\citenamefont {Haravifard}\ \emph {et~al.}(2016)\citenamefont
  {Haravifard}, \citenamefont {Graf}, \citenamefont {Feiguin}, \citenamefont
  {Batista}, \citenamefont {Lang}, \citenamefont {Silevitch}, \citenamefont
  {Srajer}, \citenamefont {Gaulin}, \citenamefont {Dabkowska},\ and\
  \citenamefont {Rosenbaum}}]{ncomms16}%
  \BibitemOpen
  \bibfield  {author} {\bibinfo {author} {\bibfnamefont {S.}~\bibnamefont
  {Haravifard}}, \bibinfo {author} {\bibfnamefont {D.}~\bibnamefont {Graf}},
  \bibinfo {author} {\bibfnamefont {A.~E.}\ \bibnamefont {Feiguin}}, \bibinfo
  {author} {\bibfnamefont {C.~D.}\ \bibnamefont {Batista}}, \bibinfo {author}
  {\bibfnamefont {J.~C.}\ \bibnamefont {Lang}}, \bibinfo {author}
  {\bibfnamefont {D.~M.}\ \bibnamefont {Silevitch}}, \bibinfo {author}
  {\bibfnamefont {G.}~\bibnamefont {Srajer}}, \bibinfo {author} {\bibfnamefont
  {B.~D.}\ \bibnamefont {Gaulin}}, \bibinfo {author} {\bibfnamefont {H.~A.}\
  \bibnamefont {Dabkowska}}, \ and\ \bibinfo {author} {\bibfnamefont {T.~F.}\
  \bibnamefont {Rosenbaum}},\ }\bibfield  {title} {\enquote {\bibinfo {title}
  {Crystallization of spin superlattices with pressure and field in the layered
  magnet {SrCu$_2$(BO$_3$)$_2$}},}\ }\href {\doibase 10.1038/ncomms11956}
  {\bibfield  {journal} {\bibinfo  {journal} {Nature Communications}\ }\textbf
  {\bibinfo {volume} {7}},\ \bibinfo {pages} {11956} (\bibinfo {year}
  {2016})}\BibitemShut {NoStop}%
\bibitem [{\citenamefont {Albrecht}\ and\ \citenamefont {Mila}(1996)}]{AM96}%
  \BibitemOpen
  \bibfield  {author} {\bibinfo {author} {\bibfnamefont {M.}~\bibnamefont
  {Albrecht}}\ and\ \bibinfo {author} {\bibfnamefont {F.}~\bibnamefont
  {Mila}},\ }\bibfield  {title} {\enquote {\bibinfo {title} {First-order
  transition between magnetic order and valence bond order in a {2D} frustrated
  {H}eisenberg model},}\ }\href {http://stacks.iop.org/0295-5075/34/i=2/a=145}
  {\bibfield  {journal} {\bibinfo  {journal} {EPL (Europhysics Letters)}\
  }\textbf {\bibinfo {volume} {34}},\ \bibinfo {pages} {145--150} (\bibinfo
  {year} {1996})}\BibitemShut {NoStop}%
\bibitem [{\citenamefont {Miyahara}\ and\ \citenamefont
  {Ueda}(1999)}]{MiUeda99}%
  \BibitemOpen
  \bibfield  {author} {\bibinfo {author} {\bibfnamefont {Shin}\ \bibnamefont
  {Miyahara}}\ and\ \bibinfo {author} {\bibfnamefont {Kazuo}\ \bibnamefont
  {Ueda}},\ }\bibfield  {title} {\enquote {\bibinfo {title} {Exact dimer ground
  state of the two dimensional {H}eisenberg spin system
  {SrCu$_2$(BO$_3$)$_2$}},}\ }\href {\doibase 10.1103/PhysRevLett.82.3701}
  {\bibfield  {journal} {\bibinfo  {journal} {Phys. Rev. Lett.}\ }\textbf
  {\bibinfo {volume} {82}},\ \bibinfo {pages} {3701--3704} (\bibinfo {year}
  {1999})}\BibitemShut {NoStop}%
\bibitem [{\citenamefont {Majumdar}\ and\ \citenamefont
  {Ghosh}(1969{\natexlab{a}})}]{MG69a}%
  \BibitemOpen
  \bibfield  {author} {\bibinfo {author} {\bibfnamefont {Chanchal~K.}\
  \bibnamefont {Majumdar}}\ and\ \bibinfo {author} {\bibfnamefont {Dipan~K.}\
  \bibnamefont {Ghosh}},\ }\bibfield  {title} {\enquote {\bibinfo {title} {On
  next-nearest-neighbor interaction in linear chain. {I}},}\ }\href {\doibase
  10.1063/1.1664978} {\bibfield  {journal} {\bibinfo  {journal} {J. Math.
  Phys.}\ }\textbf {\bibinfo {volume} {10}},\ \bibinfo {pages} {1388--1398}
  (\bibinfo {year} {1969}{\natexlab{a}})}\BibitemShut {NoStop}%
\bibitem [{\citenamefont {Majumdar}\ and\ \citenamefont
  {Ghosh}(1969{\natexlab{b}})}]{MG69b}%
  \BibitemOpen
  \bibfield  {author} {\bibinfo {author} {\bibfnamefont {Chanchal~K.}\
  \bibnamefont {Majumdar}}\ and\ \bibinfo {author} {\bibfnamefont {Dipan~K.}\
  \bibnamefont {Ghosh}},\ }\bibfield  {title} {\enquote {\bibinfo {title} {On
  next-nearest-neighbor interaction in linear chain. {II}},}\ }\href {\doibase
  10.1063/1.1664979} {\bibfield  {journal} {\bibinfo  {journal} {J. Math.
  Phys.}\ }\textbf {\bibinfo {volume} {10}},\ \bibinfo {pages} {1399--1402}
  (\bibinfo {year} {1969}{\natexlab{b}})}\BibitemShut {NoStop}%
\bibitem [{\citenamefont {Majumdar}(1970)}]{Majumdar70}%
  \BibitemOpen
  \bibfield  {author} {\bibinfo {author} {\bibfnamefont {C.~K.}\ \bibnamefont
  {Majumdar}},\ }\bibfield  {title} {\enquote {\bibinfo {title}
  {Antiferromagnetic model with known ground state},}\ }\href
  {http://stacks.iop.org/0022-3719/3/i=4/a=019} {\bibfield  {journal} {\bibinfo
   {journal} {J. Phys. C: Solid State Phys.}\ }\textbf {\bibinfo {volume}
  {3}},\ \bibinfo {pages} {911--915} (\bibinfo {year} {1970})}\BibitemShut
  {NoStop}%
\bibitem [{\citenamefont {Kageyama}\ \emph {et~al.}(2000)\citenamefont
  {Kageyama}, \citenamefont {Onizuka}, \citenamefont {Ueda}, \citenamefont
  {Nohara}, \citenamefont {Suzuki},\ and\ \citenamefont
  {Takagi}}]{Kageyama2000}%
  \BibitemOpen
  \bibfield  {author} {\bibinfo {author} {\bibfnamefont {H.}~\bibnamefont
  {Kageyama}}, \bibinfo {author} {\bibfnamefont {K.}~\bibnamefont {Onizuka}},
  \bibinfo {author} {\bibfnamefont {Y.}~\bibnamefont {Ueda}}, \bibinfo {author}
  {\bibfnamefont {M.}~\bibnamefont {Nohara}}, \bibinfo {author} {\bibfnamefont
  {H.}~\bibnamefont {Suzuki}}, \ and\ \bibinfo {author} {\bibfnamefont
  {H.}~\bibnamefont {Takagi}},\ }\bibfield  {title} {\enquote {\bibinfo {title}
  {Low-temperature specific heat study of {SrCu$_2$(BO$_3$)$_2$} with an
  exactly solvable ground state},}\ }\href {\doibase 10.1134/1.559083}
  {\bibfield  {journal} {\bibinfo  {journal} {Journal of Experimental and
  Theoretical Physics}\ }\textbf {\bibinfo {volume} {90}},\ \bibinfo {pages}
  {129--132} (\bibinfo {year} {2000})}\BibitemShut {NoStop}%
\bibitem [{\citenamefont {Miyahara}\ and\ \citenamefont
  {Ueda}(2000)}]{MiUeda00}%
  \BibitemOpen
  \bibfield  {author} {\bibinfo {author} {\bibfnamefont {S.}~\bibnamefont
  {Miyahara}}\ and\ \bibinfo {author} {\bibfnamefont {K.}~\bibnamefont
  {Ueda}},\ }\bibfield  {title} {\enquote {\bibinfo {title} {Thermodynamic
  properties of three-dimensional orthogonal dimer model for
  {SrCu$_2$(BO$_3$)$_2$}},}\ }\href@noop {} {\bibfield  {journal} {\bibinfo
  {journal} {J. Phys. Soc. Jpn. (Suppl.) B}\ }\textbf {\bibinfo {volume}
  {69}},\ \bibinfo {pages} {72--77} (\bibinfo {year} {2000})},\ \Eprint
  {http://arxiv.org/abs/cond-mat/0004260} {cond-mat/0004260} \BibitemShut
  {NoStop}%
\bibitem [{\citenamefont {Miyahara}\ and\ \citenamefont
  {Ueda}(2003)}]{MiUeda03}%
  \BibitemOpen
  \bibfield  {author} {\bibinfo {author} {\bibfnamefont {Shin}\ \bibnamefont
  {Miyahara}}\ and\ \bibinfo {author} {\bibfnamefont {Kazuo}\ \bibnamefont
  {Ueda}},\ }\bibfield  {title} {\enquote {\bibinfo {title} {Theory of the
  orthogonal dimer {H}eisenberg spin model for {SrCu$_2$(BO$_3$)$_2$}},}\
  }\href {http://stacks.iop.org/0953-8984/15/i=9/a=201} {\bibfield  {journal}
  {\bibinfo  {journal} {J. Phys.: Condens. Matter}\ }\textbf {\bibinfo {volume}
  {15}},\ \bibinfo {pages} {R327--R366} (\bibinfo {year} {2003})}\BibitemShut
  {NoStop}%
\bibitem [{Note1()}]{Note1}%
  \BibitemOpen
  \bibinfo {note} {The magnetic susceptibility, $\chi (T)$, has also been
  analyzed by series expansions \cite {MiUeda99,PhysRevB.60.6608,
  PhysRevLett.85.3958}, but these are accurate only for temperatures above the
  maximum of $\chi $.}\BibitemShut {Stop}%
\bibitem [{\citenamefont {Weihong}\ \emph {et~al.}(1999)\citenamefont
  {Weihong}, \citenamefont {Hamer},\ and\ \citenamefont
  {Oitmaa}}]{PhysRevB.60.6608}%
  \BibitemOpen
  \bibfield  {author} {\bibinfo {author} {\bibfnamefont {Zheng}\ \bibnamefont
  {Weihong}}, \bibinfo {author} {\bibfnamefont {C.~J.}\ \bibnamefont {Hamer}},
  \ and\ \bibinfo {author} {\bibfnamefont {J.}~\bibnamefont {Oitmaa}},\
  }\bibfield  {title} {\enquote {\bibinfo {title} {Series expansions for a
  {H}eisenberg antiferromagnetic model for {SrCu$_{2}$(BO$_{3}$)$_{2}$}},}\
  }\href {\doibase 10.1103/PhysRevB.60.6608} {\bibfield  {journal} {\bibinfo
  {journal} {Phys. Rev. B}\ }\textbf {\bibinfo {volume} {60}},\ \bibinfo
  {pages} {6608--6616} (\bibinfo {year} {1999})}\BibitemShut {NoStop}%
\bibitem [{\citenamefont {Knetter}\ \emph {et~al.}(2000)\citenamefont
  {Knetter}, \citenamefont {B\"uhler}, \citenamefont {M\"uller-Hartmann},\ and\
  \citenamefont {Uhrig}}]{PhysRevLett.85.3958}%
  \BibitemOpen
  \bibfield  {author} {\bibinfo {author} {\bibfnamefont {Christian}\
  \bibnamefont {Knetter}}, \bibinfo {author} {\bibfnamefont {Alexander}\
  \bibnamefont {B\"uhler}}, \bibinfo {author} {\bibfnamefont {Erwin}\
  \bibnamefont {M\"uller-Hartmann}}, \ and\ \bibinfo {author} {\bibfnamefont
  {G\"otz~S.}\ \bibnamefont {Uhrig}},\ }\bibfield  {title} {\enquote {\bibinfo
  {title} {Dispersion and symmetry of bound states in the
  {S}hastry-{S}utherland model},}\ }\href {\doibase
  10.1103/PhysRevLett.85.3958} {\bibfield  {journal} {\bibinfo  {journal}
  {Phys. Rev. Lett.}\ }\textbf {\bibinfo {volume} {85}},\ \bibinfo {pages}
  {3958--3961} (\bibinfo {year} {2000})}\BibitemShut {NoStop}%
\bibitem [{\citenamefont {Honecker}\ \emph {et~al.}(2016)\citenamefont
  {Honecker}, \citenamefont {Wessel}, \citenamefont {Kerkdyk}, \citenamefont
  {Pruschke}, \citenamefont {Mila},\ and\ \citenamefont {Normand}}]{rus1}%
  \BibitemOpen
  \bibfield  {author} {\bibinfo {author} {\bibfnamefont {A.}~\bibnamefont
  {Honecker}}, \bibinfo {author} {\bibfnamefont {S.}~\bibnamefont {Wessel}},
  \bibinfo {author} {\bibfnamefont {R.}~\bibnamefont {Kerkdyk}}, \bibinfo
  {author} {\bibfnamefont {T.}~\bibnamefont {Pruschke}}, \bibinfo {author}
  {\bibfnamefont {F.}~\bibnamefont {Mila}}, \ and\ \bibinfo {author}
  {\bibfnamefont {B.}~\bibnamefont {Normand}},\ }\bibfield  {title} {\enquote
  {\bibinfo {title} {Thermodynamic properties of highly frustrated quantum spin
  ladders: Influence of many-particle bound states},}\ }\href {\doibase
  10.1103/PhysRevB.93.054408} {\bibfield  {journal} {\bibinfo  {journal} {Phys.
  Rev. B}\ }\textbf {\bibinfo {volume} {93}},\ \bibinfo {pages} {054408}
  (\bibinfo {year} {2016})}\BibitemShut {NoStop}%
\bibitem [{\citenamefont {Alet}\ \emph {et~al.}(2016)\citenamefont {Alet},
  \citenamefont {Damle},\ and\ \citenamefont {Pujari}}]{Alet16}%
  \BibitemOpen
  \bibfield  {author} {\bibinfo {author} {\bibfnamefont {Fabien}\ \bibnamefont
  {Alet}}, \bibinfo {author} {\bibfnamefont {Kedar}\ \bibnamefont {Damle}}, \
  and\ \bibinfo {author} {\bibfnamefont {Sumiran}\ \bibnamefont {Pujari}},\
  }\bibfield  {title} {\enquote {\bibinfo {title} {Sign-problem-free {M}onte
  {C}arlo simulation of certain frustrated quantum magnets},}\ }\href {\doibase
  10.1103/PhysRevLett.117.197203} {\bibfield  {journal} {\bibinfo  {journal}
  {Phys. Rev. Lett.}\ }\textbf {\bibinfo {volume} {117}},\ \bibinfo {pages}
  {197203} (\bibinfo {year} {2016})}\BibitemShut {NoStop}%
\bibitem [{\citenamefont {Ng}\ and\ \citenamefont {Yang}(2017)}]{NgYang17}%
  \BibitemOpen
  \bibfield  {author} {\bibinfo {author} {\bibfnamefont {Kwai-Kong}\
  \bibnamefont {Ng}}\ and\ \bibinfo {author} {\bibfnamefont {Min-Fong}\
  \bibnamefont {Yang}},\ }\bibfield  {title} {\enquote {\bibinfo {title}
  {Field-induced quantum phases in a frustrated spin-dimer model: A
  sign-problem-free quantum {M}onte {C}arlo study},}\ }\href {\doibase
  10.1103/PhysRevB.95.064431} {\bibfield  {journal} {\bibinfo  {journal} {Phys.
  Rev. B}\ }\textbf {\bibinfo {volume} {95}},\ \bibinfo {pages} {064431}
  (\bibinfo {year} {2017})}\BibitemShut {NoStop}%
\bibitem [{\citenamefont {Wessel}\ \emph {et~al.}(2017)\citenamefont {Wessel},
  \citenamefont {Normand}, \citenamefont {Mila},\ and\ \citenamefont
  {Honecker}}]{rus3}%
  \BibitemOpen
  \bibfield  {author} {\bibinfo {author} {\bibfnamefont {Stefan}\ \bibnamefont
  {Wessel}}, \bibinfo {author} {\bibfnamefont {B.}~\bibnamefont {Normand}},
  \bibinfo {author} {\bibfnamefont {Fr\'ed\'eric}\ \bibnamefont {Mila}}, \ and\
  \bibinfo {author} {\bibfnamefont {Andreas}\ \bibnamefont {Honecker}},\
  }\bibfield  {title} {\enquote {\bibinfo {title} {Efficient quantum {M}onte
  {C}arlo simulations of highly frustrated magnets: the frustrated spin-1/2
  ladder},}\ }\href {\doibase 10.21468/SciPostPhys.3.1.005} {\bibfield
  {journal} {\bibinfo  {journal} {SciPost Phys.}\ }\textbf {\bibinfo {volume}
  {3}},\ \bibinfo {pages} {005} (\bibinfo {year} {2017})}\BibitemShut {NoStop}%
\bibitem [{\citenamefont {Stapmanns}\ \emph {et~al.}(2018)\citenamefont
  {Stapmanns}, \citenamefont {Corboz}, \citenamefont {Mila}, \citenamefont
  {Honecker}, \citenamefont {Normand},\ and\ \citenamefont {Wessel}}]{rus4}%
  \BibitemOpen
  \bibfield  {author} {\bibinfo {author} {\bibfnamefont {J.}~\bibnamefont
  {Stapmanns}}, \bibinfo {author} {\bibfnamefont {P.}~\bibnamefont {Corboz}},
  \bibinfo {author} {\bibfnamefont {F.}~\bibnamefont {Mila}}, \bibinfo {author}
  {\bibfnamefont {A.}~\bibnamefont {Honecker}}, \bibinfo {author}
  {\bibfnamefont {B.}~\bibnamefont {Normand}}, \ and\ \bibinfo {author}
  {\bibfnamefont {S.}~\bibnamefont {Wessel}},\ }\bibfield  {title} {\enquote
  {\bibinfo {title} {Thermal critical points and quantum critical end point in
  the frustrated bilayer {H}eisenberg antiferromagnet},}\ }\href {\doibase
  10.1103/PhysRevLett.121.127201} {\bibfield  {journal} {\bibinfo  {journal}
  {Phys. Rev. Lett.}\ }\textbf {\bibinfo {volume} {121}},\ \bibinfo {pages}
  {127201} (\bibinfo {year} {2018})}\BibitemShut {NoStop}%
\bibitem [{\citenamefont {Lin}\ and\ \citenamefont {Shen}(2000)}]{LinShen00}%
  \BibitemOpen
  \bibfield  {author} {\bibinfo {author} {\bibfnamefont {Hai-Qing}\
  \bibnamefont {Lin}}\ and\ \bibinfo {author} {\bibfnamefont {J.~L.}\
  \bibnamefont {Shen}},\ }\bibfield  {title} {\enquote {\bibinfo {title} {Exact
  ground states and excited states of net spin models},}\ }\href {\doibase
  10.1143/JPSJ.69.878} {\bibfield  {journal} {\bibinfo  {journal} {J. Phys.
  Soc. Jpn.}\ }\textbf {\bibinfo {volume} {69}},\ \bibinfo {pages} {878--882}
  (\bibinfo {year} {2000})}\BibitemShut {NoStop}%
\bibitem [{\citenamefont {Lin}\ \emph {et~al.}(2002)\citenamefont {Lin},
  \citenamefont {Shen},\ and\ \citenamefont {Shik}}]{LinShik02}%
  \BibitemOpen
  \bibfield  {author} {\bibinfo {author} {\bibfnamefont {H.~Q.}\ \bibnamefont
  {Lin}}, \bibinfo {author} {\bibfnamefont {J.~L.}\ \bibnamefont {Shen}}, \
  and\ \bibinfo {author} {\bibfnamefont {H.~Y.}\ \bibnamefont {Shik}},\
  }\bibfield  {title} {\enquote {\bibinfo {title} {Exactly soluble quantum spin
  models on a double layer: The net spin model},}\ }\href {\doibase
  10.1103/PhysRevB.66.184402} {\bibfield  {journal} {\bibinfo  {journal} {Phys.
  Rev. B}\ }\textbf {\bibinfo {volume} {66}},\ \bibinfo {pages} {184402}
  (\bibinfo {year} {2002})}\BibitemShut {NoStop}%
\bibitem [{\citenamefont {Richter}\ \emph {et~al.}(2006)\citenamefont
  {Richter}, \citenamefont {Derzhko},\ and\ \citenamefont
  {Krokhmalskii}}]{RDK06}%
  \BibitemOpen
  \bibfield  {author} {\bibinfo {author} {\bibfnamefont {Johannes}\
  \bibnamefont {Richter}}, \bibinfo {author} {\bibfnamefont {Oleg}\
  \bibnamefont {Derzhko}}, \ and\ \bibinfo {author} {\bibfnamefont {Taras}\
  \bibnamefont {Krokhmalskii}},\ }\bibfield  {title} {\enquote {\bibinfo
  {title} {Finite-temperature order-disorder phase transition in a frustrated
  bilayer quantum {H}eisenberg antiferromagnet in strong magnetic fields},}\
  }\href {\doibase 10.1103/PhysRevB.74.144430} {\bibfield  {journal} {\bibinfo
  {journal} {Phys. Rev. B}\ }\textbf {\bibinfo {volume} {74}},\ \bibinfo
  {pages} {144430} (\bibinfo {year} {2006})}\BibitemShut {NoStop}%
\bibitem [{\citenamefont {Derzhko}\ \emph {et~al.}(2007)\citenamefont
  {Derzhko}, \citenamefont {Richter}, \citenamefont {Honecker},\ and\
  \citenamefont {Schmidt}}]{DRHS07}%
  \BibitemOpen
  \bibfield  {author} {\bibinfo {author} {\bibfnamefont {O.}~\bibnamefont
  {Derzhko}}, \bibinfo {author} {\bibfnamefont {J.}~\bibnamefont {Richter}},
  \bibinfo {author} {\bibfnamefont {A.}~\bibnamefont {Honecker}}, \ and\
  \bibinfo {author} {\bibfnamefont {H.-J.}\ \bibnamefont {Schmidt}},\
  }\bibfield  {title} {\enquote {\bibinfo {title} {Universal properties of
  highly frustrated quantum magnets in strong magnetic fields},}\ }\href
  {\doibase 10.1063/1.2780166} {\bibfield  {journal} {\bibinfo  {journal} {Low
  Temp. Phys.}\ }\textbf {\bibinfo {volume} {33}},\ \bibinfo {pages} {745--756}
  (\bibinfo {year} {2007})}\BibitemShut {NoStop}%
\bibitem [{\citenamefont {Derzhko}\ \emph {et~al.}(2010)\citenamefont
  {Derzhko}, \citenamefont {Krokhmalskii},\ and\ \citenamefont
  {Richter}}]{DKR10}%
  \BibitemOpen
  \bibfield  {author} {\bibinfo {author} {\bibfnamefont {Oleg}\ \bibnamefont
  {Derzhko}}, \bibinfo {author} {\bibfnamefont {Taras}\ \bibnamefont
  {Krokhmalskii}}, \ and\ \bibinfo {author} {\bibfnamefont {Johannes}\
  \bibnamefont {Richter}},\ }\bibfield  {title} {\enquote {\bibinfo {title}
  {Emergent {I}sing degrees of freedom in frustrated two-leg ladder and bilayer
  $s = 1/2$ {H}eisenberg antiferromagnets},}\ }\href {\doibase
  10.1103/PhysRevB.82.214412} {\bibfield  {journal} {\bibinfo  {journal} {Phys.
  Rev. B}\ }\textbf {\bibinfo {volume} {82}},\ \bibinfo {pages} {214412}
  (\bibinfo {year} {2010})}\BibitemShut {NoStop}%
\bibitem [{\citenamefont {M\"uller-Hartmann}\ \emph {et~al.}(2000)\citenamefont
  {M\"uller-Hartmann}, \citenamefont {Singh}, \citenamefont {Knetter},\ and\
  \citenamefont {Uhrig}}]{PhysRevLett.84.1808}%
  \BibitemOpen
  \bibfield  {author} {\bibinfo {author} {\bibfnamefont {Erwin}\ \bibnamefont
  {M\"uller-Hartmann}}, \bibinfo {author} {\bibfnamefont {Rajiv R.~P.}\
  \bibnamefont {Singh}}, \bibinfo {author} {\bibfnamefont {Christian}\
  \bibnamefont {Knetter}}, \ and\ \bibinfo {author} {\bibfnamefont {G\"otz~S.}\
  \bibnamefont {Uhrig}},\ }\bibfield  {title} {\enquote {\bibinfo {title}
  {Exact demonstration of magnetization plateaus and first-order dimer-{N}\'eel
  phase transitions in a modified {S}hastry-{S}utherland model for
  {SrCu$_{2}$(BO$_{3}$)$_{2}$}},}\ }\href {\doibase
  10.1103/PhysRevLett.84.1808} {\bibfield  {journal} {\bibinfo  {journal}
  {Phys. Rev. Lett.}\ }\textbf {\bibinfo {volume} {84}},\ \bibinfo {pages}
  {1808--1811} (\bibinfo {year} {2000})}\BibitemShut {NoStop}%
\bibitem [{\citenamefont {Corboz}\ and\ \citenamefont
  {Mila}(2013)}]{CorbozMila13}%
  \BibitemOpen
  \bibfield  {author} {\bibinfo {author} {\bibfnamefont {Philippe}\
  \bibnamefont {Corboz}}\ and\ \bibinfo {author} {\bibfnamefont {Fr\'ed\'eric}\
  \bibnamefont {Mila}},\ }\bibfield  {title} {\enquote {\bibinfo {title}
  {Tensor network study of the {S}hastry-{S}utherland model in zero magnetic
  field},}\ }\href {\doibase 10.1103/PhysRevB.87.115144} {\bibfield  {journal}
  {\bibinfo  {journal} {Phys. Rev. B}\ }\textbf {\bibinfo {volume} {87}},\
  \bibinfo {pages} {115144} (\bibinfo {year} {2013})}\BibitemShut {NoStop}%
\bibitem [{\citenamefont {Koga}\ and\ \citenamefont {Kawakami}(2000)}]{rkk}%
  \BibitemOpen
  \bibfield  {author} {\bibinfo {author} {\bibfnamefont {Akihisa}\ \bibnamefont
  {Koga}}\ and\ \bibinfo {author} {\bibfnamefont {Norio}\ \bibnamefont
  {Kawakami}},\ }\bibfield  {title} {\enquote {\bibinfo {title} {Quantum phase
  transitions in the {S}hastry-{S}utherland model for
  {SrCu$_2$(BO$_3$)$_2$}},}\ }\href {\doibase 10.1103/PhysRevLett.84.4461}
  {\bibfield  {journal} {\bibinfo  {journal} {Phys. Rev. Lett.}\ }\textbf
  {\bibinfo {volume} {84}},\ \bibinfo {pages} {4461--4464} (\bibinfo {year}
  {2000})}\BibitemShut {NoStop}%
\bibitem [{\citenamefont {Takushima}\ \emph {et~al.}(2001)\citenamefont
  {Takushima}, \citenamefont {Koga},\ and\ \citenamefont {Kawakami}}]{rtkk}%
  \BibitemOpen
  \bibfield  {author} {\bibinfo {author} {\bibfnamefont {Yoshihiro}\
  \bibnamefont {Takushima}}, \bibinfo {author} {\bibfnamefont {Akihisa}\
  \bibnamefont {Koga}}, \ and\ \bibinfo {author} {\bibfnamefont {Norio}\
  \bibnamefont {Kawakami}},\ }\bibfield  {title} {\enquote {\bibinfo {title}
  {Competing spin-gap phases in a frustrated quantum spin system in two
  dimensions},}\ }\href {\doibase 10.1143/JPSJ.70.1369} {\bibfield  {journal}
  {\bibinfo  {journal} {J. Phys. Soc. Jpn.}\ }\textbf {\bibinfo {volume}
  {70}},\ \bibinfo {pages} {1369--1374} (\bibinfo {year} {2001})}\BibitemShut
  {NoStop}%
\bibitem [{\citenamefont {L\"auchli}\ \emph {et~al.}(2002)\citenamefont
  {L\"auchli}, \citenamefont {Wessel},\ and\ \citenamefont {Sigrist}}]{LWS02}%
  \BibitemOpen
  \bibfield  {author} {\bibinfo {author} {\bibfnamefont {Andreas}\ \bibnamefont
  {L\"auchli}}, \bibinfo {author} {\bibfnamefont {Stefan}\ \bibnamefont
  {Wessel}}, \ and\ \bibinfo {author} {\bibfnamefont {Manfred}\ \bibnamefont
  {Sigrist}},\ }\bibfield  {title} {\enquote {\bibinfo {title} {Phase diagram
  of the quadrumerized {S}hastry-{S}utherland model},}\ }\href {\doibase
  10.1103/PhysRevB.66.014401} {\bibfield  {journal} {\bibinfo  {journal} {Phys.
  Rev. B}\ }\textbf {\bibinfo {volume} {66}},\ \bibinfo {pages} {014401}
  (\bibinfo {year} {2002})}\BibitemShut {NoStop}%
\bibitem [{\citenamefont {Sandvik}(1999)}]{sse0}%
  \BibitemOpen
  \bibfield  {author} {\bibinfo {author} {\bibfnamefont {Anders~W.}\
  \bibnamefont {Sandvik}},\ }\bibfield  {title} {\enquote {\bibinfo {title}
  {Stochastic series expansion method with operator-loop update},}\ }\href
  {\doibase 10.1103/PhysRevB.59.R14157} {\bibfield  {journal} {\bibinfo
  {journal} {Phys. Rev. B}\ }\textbf {\bibinfo {volume} {59}},\ \bibinfo
  {pages} {R14157--R14160} (\bibinfo {year} {1999})}\BibitemShut {NoStop}%
\bibitem [{\citenamefont {Sylju\aa{}sen}\ and\ \citenamefont
  {Sandvik}(2002)}]{sse1}%
  \BibitemOpen
  \bibfield  {author} {\bibinfo {author} {\bibfnamefont {Olav~F.}\ \bibnamefont
  {Sylju\aa{}sen}}\ and\ \bibinfo {author} {\bibfnamefont {Anders~W.}\
  \bibnamefont {Sandvik}},\ }\bibfield  {title} {\enquote {\bibinfo {title}
  {{Q}uantum {M}onte {C}arlo with directed loops},}\ }\href {\doibase
  10.1103/PhysRevE.66.046701} {\bibfield  {journal} {\bibinfo  {journal} {Phys.
  Rev. E}\ }\textbf {\bibinfo {volume} {66}},\ \bibinfo {pages} {046701}
  (\bibinfo {year} {2002})}\BibitemShut {NoStop}%
\bibitem [{\citenamefont {Alet}\ \emph {et~al.}(2005)\citenamefont {Alet},
  \citenamefont {Wessel},\ and\ \citenamefont {Troyer}}]{sse2}%
  \BibitemOpen
  \bibfield  {author} {\bibinfo {author} {\bibfnamefont {Fabien}\ \bibnamefont
  {Alet}}, \bibinfo {author} {\bibfnamefont {Stefan}\ \bibnamefont {Wessel}}, \
  and\ \bibinfo {author} {\bibfnamefont {Matthias}\ \bibnamefont {Troyer}},\
  }\bibfield  {title} {\enquote {\bibinfo {title} {Generalized directed loop
  method for quantum {M}onte {C}arlo simulations},}\ }\href {\doibase
  10.1103/PhysRevE.71.036706} {\bibfield  {journal} {\bibinfo  {journal} {Phys.
  Rev. E}\ }\textbf {\bibinfo {volume} {71}},\ \bibinfo {pages} {036706}
  (\bibinfo {year} {2005})}\BibitemShut {NoStop}%
\bibitem [{\citenamefont {Evertz}(2003)}]{EvertzLoop03}%
  \BibitemOpen
  \bibfield  {author} {\bibinfo {author} {\bibfnamefont {H.~G.}\ \bibnamefont
  {Evertz}},\ }\bibfield  {title} {\enquote {\bibinfo {title} {The loop
  algorithm},}\ }\href {\doibase 10.1080/0001873021000049195} {\bibfield
  {journal} {\bibinfo  {journal} {Advances in Physics}\ }\textbf {\bibinfo
  {volume} {52}},\ \bibinfo {pages} {1--66} (\bibinfo {year}
  {2003})}\BibitemShut {NoStop}%
\bibitem [{\citenamefont {Troyer}\ and\ \citenamefont
  {Wiese}(2005)}]{PhysRevLett.94.170201}%
  \BibitemOpen
  \bibfield  {author} {\bibinfo {author} {\bibfnamefont {Matthias}\
  \bibnamefont {Troyer}}\ and\ \bibinfo {author} {\bibfnamefont {Uwe-Jens}\
  \bibnamefont {Wiese}},\ }\bibfield  {title} {\enquote {\bibinfo {title}
  {Computational complexity and fundamental limitations to fermionic quantum
  {Monte Carlo} simulations},}\ }\href {\doibase 10.1103/PhysRevLett.94.170201}
  {\bibfield  {journal} {\bibinfo  {journal} {Phys. Rev. Lett.}\ }\textbf
  {\bibinfo {volume} {94}},\ \bibinfo {pages} {170201} (\bibinfo {year}
  {2005})}\BibitemShut {NoStop}%
\bibitem [{\citenamefont {{Bleaney, F. R. S.}}\ and\ \citenamefont
  {Bowers}(1952)}]{Bleaney451}%
  \BibitemOpen
  \bibfield  {author} {\bibinfo {author} {\bibfnamefont {B.}~\bibnamefont
  {{Bleaney, F. R. S.}}}\ and\ \bibinfo {author} {\bibfnamefont {K.~D.}\
  \bibnamefont {Bowers}},\ }\bibfield  {title} {\enquote {\bibinfo {title}
  {Anomalous paramagnetism of copper acetate},}\ }\href {\doibase
  10.1098/rspa.1952.0181} {\bibfield  {journal} {\bibinfo  {journal} {Proc. R.
  Soc. London, Ser. A}\ }\textbf {\bibinfo {volume} {214}},\ \bibinfo {pages}
  {451--465} (\bibinfo {year} {1952})}\BibitemShut {NoStop}%
\bibitem [{\citenamefont {Johnston}\ \emph {et~al.}(2000)\citenamefont
  {Johnston}, \citenamefont {Kremer}, \citenamefont {Troyer}, \citenamefont
  {Wang}, \citenamefont {Kl\"umper}, \citenamefont {Bud'ko}, \citenamefont
  {Panchula},\ and\ \citenamefont {Canfield}}]{PhysRevB.61.9558}%
  \BibitemOpen
  \bibfield  {author} {\bibinfo {author} {\bibfnamefont {D.~C.}\ \bibnamefont
  {Johnston}}, \bibinfo {author} {\bibfnamefont {R.~K.}\ \bibnamefont
  {Kremer}}, \bibinfo {author} {\bibfnamefont {M.}~\bibnamefont {Troyer}},
  \bibinfo {author} {\bibfnamefont {X.}~\bibnamefont {Wang}}, \bibinfo {author}
  {\bibfnamefont {A.}~\bibnamefont {Kl\"umper}}, \bibinfo {author}
  {\bibfnamefont {S.~L.}\ \bibnamefont {Bud'ko}}, \bibinfo {author}
  {\bibfnamefont {A.~F.}\ \bibnamefont {Panchula}}, \ and\ \bibinfo {author}
  {\bibfnamefont {P.~C.}\ \bibnamefont {Canfield}},\ }\bibfield  {title}
  {\enquote {\bibinfo {title} {Thermodynamics of spin {$S=1/2$}
  antiferromagnetic uniform and alternating-exchange {H}eisenberg chains},}\
  }\href {\doibase 10.1103/PhysRevB.61.9558} {\bibfield  {journal} {\bibinfo
  {journal} {Phys. Rev. B}\ }\textbf {\bibinfo {volume} {61}},\ \bibinfo
  {pages} {9558--9606} (\bibinfo {year} {2000})}\BibitemShut {NoStop}%
\bibitem [{\citenamefont {Deisenhofer}\ \emph {et~al.}(2006)\citenamefont
  {Deisenhofer}, \citenamefont {Eremina}, \citenamefont {Pimenov},
  \citenamefont {Gavrilova}, \citenamefont {Berger}, \citenamefont {Johnsson},
  \citenamefont {Lemmens}, \citenamefont {Krug~von Nidda}, \citenamefont
  {Loidl}, \citenamefont {Lee},\ and\ \citenamefont
  {Whangbo}}]{PhysRevB.74.174421}%
  \BibitemOpen
  \bibfield  {author} {\bibinfo {author} {\bibfnamefont {J.}~\bibnamefont
  {Deisenhofer}}, \bibinfo {author} {\bibfnamefont {R.~M.}\ \bibnamefont
  {Eremina}}, \bibinfo {author} {\bibfnamefont {A.}~\bibnamefont {Pimenov}},
  \bibinfo {author} {\bibfnamefont {T.}~\bibnamefont {Gavrilova}}, \bibinfo
  {author} {\bibfnamefont {H.}~\bibnamefont {Berger}}, \bibinfo {author}
  {\bibfnamefont {M.}~\bibnamefont {Johnsson}}, \bibinfo {author}
  {\bibfnamefont {P.}~\bibnamefont {Lemmens}}, \bibinfo {author} {\bibfnamefont
  {H.-A.}\ \bibnamefont {Krug~von Nidda}}, \bibinfo {author} {\bibfnamefont
  {A.}~\bibnamefont {Loidl}}, \bibinfo {author} {\bibfnamefont {K.-S.}\
  \bibnamefont {Lee}}, \ and\ \bibinfo {author} {\bibfnamefont {M.-H.}\
  \bibnamefont {Whangbo}},\ }\bibfield  {title} {\enquote {\bibinfo {title}
  {Structural and magnetic dimers in the spin-gapped system
  {Cu}{Te}$_{2}${O}$_{5}$},}\ }\href {\doibase 10.1103/PhysRevB.74.174421}
  {\bibfield  {journal} {\bibinfo  {journal} {Phys. Rev. B}\ }\textbf {\bibinfo
  {volume} {74}},\ \bibinfo {pages} {174421} (\bibinfo {year}
  {2006})}\BibitemShut {NoStop}%
\bibitem [{\citenamefont {Bernu}\ and\ \citenamefont
  {Misguich}(2001)}]{PhysRevB.63.134409}%
  \BibitemOpen
  \bibfield  {author} {\bibinfo {author} {\bibfnamefont {B.}~\bibnamefont
  {Bernu}}\ and\ \bibinfo {author} {\bibfnamefont {G.}~\bibnamefont
  {Misguich}},\ }\bibfield  {title} {\enquote {\bibinfo {title} {Specific heat
  and high-temperature series of lattice models: Interpolation scheme and
  examples on quantum spin systems in one and two dimensions},}\ }\href
  {\doibase 10.1103/PhysRevB.63.134409} {\bibfield  {journal} {\bibinfo
  {journal} {Phys. Rev. B}\ }\textbf {\bibinfo {volume} {63}},\ \bibinfo
  {pages} {134409} (\bibinfo {year} {2001})}\BibitemShut {NoStop}%
\bibitem [{\citenamefont {Bernu}\ and\ \citenamefont
  {Lhuillier}(2015)}]{PhysRevLett.114.057201}%
  \BibitemOpen
  \bibfield  {author} {\bibinfo {author} {\bibfnamefont {B.}~\bibnamefont
  {Bernu}}\ and\ \bibinfo {author} {\bibfnamefont {C.}~\bibnamefont
  {Lhuillier}},\ }\bibfield  {title} {\enquote {\bibinfo {title} {Spin
  susceptibility of quantum magnets from high to low temperatures},}\ }\href
  {\doibase 10.1103/PhysRevLett.114.057201} {\bibfield  {journal} {\bibinfo
  {journal} {Phys. Rev. Lett.}\ }\textbf {\bibinfo {volume} {114}},\ \bibinfo
  {pages} {057201} (\bibinfo {year} {2015})}\BibitemShut {NoStop}%
\bibitem [{\citenamefont {Schmidt}\ \emph {et~al.}(2017)\citenamefont
  {Schmidt}, \citenamefont {Hauser}, \citenamefont {Lohmann},\ and\
  \citenamefont {Richter}}]{PhysRevE.95.042110}%
  \BibitemOpen
  \bibfield  {author} {\bibinfo {author} {\bibfnamefont {Heinz-J\"urgen}\
  \bibnamefont {Schmidt}}, \bibinfo {author} {\bibfnamefont {Andreas}\
  \bibnamefont {Hauser}}, \bibinfo {author} {\bibfnamefont {Andre}\
  \bibnamefont {Lohmann}}, \ and\ \bibinfo {author} {\bibfnamefont {Johannes}\
  \bibnamefont {Richter}},\ }\bibfield  {title} {\enquote {\bibinfo {title}
  {Interpolation between low and high temperatures of the specific heat for
  spin systems},}\ }\href {\doibase 10.1103/PhysRevE.95.042110} {\bibfield
  {journal} {\bibinfo  {journal} {Phys. Rev. E}\ }\textbf {\bibinfo {volume}
  {95}},\ \bibinfo {pages} {042110} (\bibinfo {year} {2017})}\BibitemShut
  {NoStop}%
\bibitem [{\citenamefont {Fukumoto}(2000)}]{Fukomoto00}%
  \BibitemOpen
  \bibfield  {author} {\bibinfo {author} {\bibfnamefont {Yoshiyuki}\
  \bibnamefont {Fukumoto}},\ }\bibfield  {title} {\enquote {\bibinfo {title}
  {Two-triplet-dimer excitation spectra in the {S}hastry-{S}utherland model for
  {Sr}{Cu}$_2$({BO}$_3$)$_2$},}\ }\href {\doibase 10.1143/JPSJ.69.2755}
  {\bibfield  {journal} {\bibinfo  {journal} {J. Phys. Soc. Jpn.}\ }\textbf
  {\bibinfo {volume} {69}},\ \bibinfo {pages} {2755--2758} (\bibinfo {year}
  {2000})}\BibitemShut {NoStop}%
\bibitem [{\citenamefont {Totsuka}\ \emph {et~al.}(2001)\citenamefont
  {Totsuka}, \citenamefont {Miyahara},\ and\ \citenamefont
  {Ueda}}]{PhysRevLett.86.520}%
  \BibitemOpen
  \bibfield  {author} {\bibinfo {author} {\bibfnamefont {K.}~\bibnamefont
  {Totsuka}}, \bibinfo {author} {\bibfnamefont {S.}~\bibnamefont {Miyahara}}, \
  and\ \bibinfo {author} {\bibfnamefont {K.}~\bibnamefont {Ueda}},\ }\bibfield
  {title} {\enquote {\bibinfo {title} {Low-lying magnetic excitation of the
  {S}hastry-{S}utherland model},}\ }\href {\doibase 10.1103/PhysRevLett.86.520}
  {\bibfield  {journal} {\bibinfo  {journal} {Phys. Rev. Lett.}\ }\textbf
  {\bibinfo {volume} {86}},\ \bibinfo {pages} {520--523} (\bibinfo {year}
  {2001})}\BibitemShut {NoStop}%
\bibitem [{\citenamefont {Schnack}\ \emph {et~al.}(2018)\citenamefont
  {Schnack}, \citenamefont {Schulenburg},\ and\ \citenamefont
  {Richter}}]{SSR18}%
  \BibitemOpen
  \bibfield  {author} {\bibinfo {author} {\bibfnamefont {J\"urgen}\
  \bibnamefont {Schnack}}, \bibinfo {author} {\bibfnamefont {J\"org}\
  \bibnamefont {Schulenburg}}, \ and\ \bibinfo {author} {\bibfnamefont
  {Johannes}\ \bibnamefont {Richter}},\ }\bibfield  {title} {\enquote {\bibinfo
  {title} {Magnetism of the {$N=42$} kagome lattice antiferromagnet},}\ }\href
  {\doibase 10.1103/PhysRevB.98.094423} {\bibfield  {journal} {\bibinfo
  {journal} {Phys. Rev. B}\ }\textbf {\bibinfo {volume} {98}},\ \bibinfo
  {pages} {094423} (\bibinfo {year} {2018})}\BibitemShut {NoStop}%
\bibitem [{\citenamefont {Verstraete}\ and\ \citenamefont
  {Cirac}(2004)}]{tnref-verstraete04}%
  \BibitemOpen
  \bibfield  {author} {\bibinfo {author} {\bibfnamefont {F.}~\bibnamefont
  {Verstraete}}\ and\ \bibinfo {author} {\bibfnamefont {J.~I.}\ \bibnamefont
  {Cirac}},\ }\bibfield  {title} {\enquote {\bibinfo {title} {Renormalization
  algorithms for quantum-many body systems in two and higher dimensions},}\
  }\href {http://arxiv.org/abs/cond-mat/0407066} {\bibfield  {journal}
  {\bibinfo  {journal} {arXiv:cond-mat/0407066}\ } (\bibinfo {year}
  {2004})}\BibitemShut {NoStop}%
\bibitem [{\citenamefont {Nishio}\ \emph {et~al.}(2004)\citenamefont {Nishio},
  \citenamefont {Maeshima}, \citenamefont {Gendiar},\ and\ \citenamefont
  {Nishino}}]{tnref-nishio04}%
  \BibitemOpen
  \bibfield  {author} {\bibinfo {author} {\bibfnamefont {Y.}~\bibnamefont
  {Nishio}}, \bibinfo {author} {\bibfnamefont {N.}~\bibnamefont {Maeshima}},
  \bibinfo {author} {\bibfnamefont {A.}~\bibnamefont {Gendiar}}, \ and\
  \bibinfo {author} {\bibfnamefont {T.}~\bibnamefont {Nishino}},\ }\bibfield
  {title} {\enquote {\bibinfo {title} {Tensor product variational formulation
  for quantum systems},}\ }\href {http://arxiv.org/abs/cond-mat/0401115}
  {\bibfield  {journal} {\bibinfo  {journal} {arXiv:cond-mat/0401115}\ }
  (\bibinfo {year} {2004})}\BibitemShut {NoStop}%
\bibitem [{\citenamefont {Jordan}\ \emph {et~al.}(2008)\citenamefont {Jordan},
  \citenamefont {Or{\'u}s}, \citenamefont {Vidal}, \citenamefont {Verstraete},\
  and\ \citenamefont {Cirac}}]{tnref-jordan08}%
  \BibitemOpen
  \bibfield  {author} {\bibinfo {author} {\bibfnamefont {J.}~\bibnamefont
  {Jordan}}, \bibinfo {author} {\bibfnamefont {R.}~\bibnamefont {Or{\'u}s}},
  \bibinfo {author} {\bibfnamefont {G.}~\bibnamefont {Vidal}}, \bibinfo
  {author} {\bibfnamefont {F.}~\bibnamefont {Verstraete}}, \ and\ \bibinfo
  {author} {\bibfnamefont {J.~I.}\ \bibnamefont {Cirac}},\ }\bibfield  {title}
  {\enquote {\bibinfo {title} {Classical simulation of infinite-size quantum
  lattice systems in two spatial dimensions},}\ }\href {\doibase
  10.1103/PhysRevLett.101.250602} {\bibfield  {journal} {\bibinfo  {journal}
  {Phys. Rev. Lett.}\ }\textbf {\bibinfo {volume} {101}},\ \bibinfo {pages}
  {250602} (\bibinfo {year} {2008})}\BibitemShut {NoStop}%
\bibitem [{\citenamefont {Corboz}\ \emph {et~al.}(2014)\citenamefont {Corboz},
  \citenamefont {Rice},\ and\ \citenamefont {Troyer}}]{tnref-corboz14}%
  \BibitemOpen
  \bibfield  {author} {\bibinfo {author} {\bibfnamefont {Philippe}\
  \bibnamefont {Corboz}}, \bibinfo {author} {\bibfnamefont {T.~M.}\
  \bibnamefont {Rice}}, \ and\ \bibinfo {author} {\bibfnamefont {Matthias}\
  \bibnamefont {Troyer}},\ }\bibfield  {title} {\enquote {\bibinfo {title}
  {Competing states in the {$t$-$J$} model: Uniform $d$-wave state versus
  stripe state},}\ }\href {\doibase 10.1103/PhysRevLett.113.046402} {\bibfield
  {journal} {\bibinfo  {journal} {Phys. Rev. Lett.}\ }\textbf {\bibinfo
  {volume} {113}},\ \bibinfo {pages} {046402} (\bibinfo {year}
  {2014})}\BibitemShut {NoStop}%
\bibitem [{\citenamefont {Nishino}\ and\ \citenamefont
  {Okunishi}(1996)}]{tnref-nishino96}%
  \BibitemOpen
  \bibfield  {author} {\bibinfo {author} {\bibfnamefont {Tomotoshi}\
  \bibnamefont {Nishino}}\ and\ \bibinfo {author} {\bibfnamefont {Kouichi}\
  \bibnamefont {Okunishi}},\ }\bibfield  {title} {\enquote {\bibinfo {title}
  {Corner transfer matrix renormalization group method},}\ }\href {\doibase
  10.1143/JPSJ.65.891} {\bibfield  {journal} {\bibinfo  {journal} {J. Phys.
  Soc. Jpn.}\ }\textbf {\bibinfo {volume} {65}},\ \bibinfo {pages} {891--894}
  (\bibinfo {year} {1996})}\BibitemShut {NoStop}%
\bibitem [{\citenamefont {Or{\'u}s}\ and\ \citenamefont
  {Vidal}(2009)}]{tnref-orus09}%
  \BibitemOpen
  \bibfield  {author} {\bibinfo {author} {\bibfnamefont {Rom\'an}\ \bibnamefont
  {Or{\'u}s}}\ and\ \bibinfo {author} {\bibfnamefont {Guifr{\'e}}\ \bibnamefont
  {Vidal}},\ }\bibfield  {title} {\enquote {\bibinfo {title} {Simulation of
  two-dimensional quantum systems on an infinite lattice revisited: {Corner}
  transfer matrix for tensor contraction},}\ }\href {\doibase
  10.1103/PhysRevB.80.094403} {\bibfield  {journal} {\bibinfo  {journal} {Phys.
  Rev. B}\ }\textbf {\bibinfo {volume} {80}},\ \bibinfo {pages} {094403}
  (\bibinfo {year} {2009})}\BibitemShut {NoStop}%
\bibitem [{\citenamefont {Singh}\ \emph {et~al.}(2011)\citenamefont {Singh},
  \citenamefont {Pfeifer},\ and\ \citenamefont {Vidal}}]{tnref-singh11}%
  \BibitemOpen
  \bibfield  {author} {\bibinfo {author} {\bibfnamefont {Sukhwinder}\
  \bibnamefont {Singh}}, \bibinfo {author} {\bibfnamefont {Robert N.~C.}\
  \bibnamefont {Pfeifer}}, \ and\ \bibinfo {author} {\bibfnamefont {Guifre}\
  \bibnamefont {Vidal}},\ }\bibfield  {title} {\enquote {\bibinfo {title}
  {Tensor network states and algorithms in the presence of a global {U}(1)
  symmetry},}\ }\href {\doibase 10.1103/PhysRevB.83.115125} {\bibfield
  {journal} {\bibinfo  {journal} {Phys. Rev. B}\ }\textbf {\bibinfo {volume}
  {83}},\ \bibinfo {pages} {115125} (\bibinfo {year} {2011})}\BibitemShut
  {NoStop}%
\bibitem [{\citenamefont {Bauer}\ \emph {et~al.}(2011)\citenamefont {Bauer},
  \citenamefont {Corboz}, \citenamefont {Or{\'u}s},\ and\ \citenamefont
  {Troyer}}]{tnref-bauer11}%
  \BibitemOpen
  \bibfield  {author} {\bibinfo {author} {\bibfnamefont {B.}~\bibnamefont
  {Bauer}}, \bibinfo {author} {\bibfnamefont {P.}~\bibnamefont {Corboz}},
  \bibinfo {author} {\bibfnamefont {R.}~\bibnamefont {Or{\'u}s}}, \ and\
  \bibinfo {author} {\bibfnamefont {M.}~\bibnamefont {Troyer}},\ }\bibfield
  {title} {\enquote {\bibinfo {title} {Implementing global {Abelian} symmetries
  in projected entangled-pair state algorithms},}\ }\href {\doibase
  10.1103/PhysRevB.83.125106} {\bibfield  {journal} {\bibinfo  {journal} {Phys.
  Rev. B}\ }\textbf {\bibinfo {volume} {83}},\ \bibinfo {pages} {125106}
  (\bibinfo {year} {2011})}\BibitemShut {NoStop}%
\bibitem [{\citenamefont {Corboz}(2016{\natexlab{a}})}]{tnref-corboz16b}%
  \BibitemOpen
  \bibfield  {author} {\bibinfo {author} {\bibfnamefont {Philippe}\
  \bibnamefont {Corboz}},\ }\bibfield  {title} {\enquote {\bibinfo {title}
  {Variational optimization with infinite projected entangled-pair states},}\
  }\href {\doibase 10.1103/PhysRevB.94.035133} {\bibfield  {journal} {\bibinfo
  {journal} {Phys. Rev. B}\ }\textbf {\bibinfo {volume} {94}},\ \bibinfo
  {pages} {035133} (\bibinfo {year} {2016}{\natexlab{a}})}\BibitemShut
  {NoStop}%
\bibitem [{\citenamefont {Jiang}\ \emph {et~al.}(2008)\citenamefont {Jiang},
  \citenamefont {Weng},\ and\ \citenamefont {Xiang}}]{tnref-jiang08}%
  \BibitemOpen
  \bibfield  {author} {\bibinfo {author} {\bibfnamefont {H.~C.}\ \bibnamefont
  {Jiang}}, \bibinfo {author} {\bibfnamefont {Z.~Y.}\ \bibnamefont {Weng}}, \
  and\ \bibinfo {author} {\bibfnamefont {T.}~\bibnamefont {Xiang}},\ }\bibfield
   {title} {\enquote {\bibinfo {title} {Accurate determination of tensor
  network state of quantum lattice models in two dimensions},}\ }\href
  {\doibase 10.1103/PhysRevLett.101.090603} {\bibfield  {journal} {\bibinfo
  {journal} {Phys. Rev. Lett.}\ }\textbf {\bibinfo {volume} {101}},\ \bibinfo
  {pages} {090603} (\bibinfo {year} {2008})}\BibitemShut {NoStop}%
\bibitem [{\citenamefont {Corboz}\ \emph {et~al.}(2010)\citenamefont {Corboz},
  \citenamefont {Or{\'u}s}, \citenamefont {Bauer},\ and\ \citenamefont
  {Vidal}}]{tnref-corboz10}%
  \BibitemOpen
  \bibfield  {author} {\bibinfo {author} {\bibfnamefont {Philippe}\
  \bibnamefont {Corboz}}, \bibinfo {author} {\bibfnamefont {Rom\'an}\
  \bibnamefont {Or{\'u}s}}, \bibinfo {author} {\bibfnamefont {Bela}\
  \bibnamefont {Bauer}}, \ and\ \bibinfo {author} {\bibfnamefont {Guifr{\'e}}\
  \bibnamefont {Vidal}},\ }\bibfield  {title} {\enquote {\bibinfo {title}
  {Simulation of strongly correlated fermions in two spatial dimensions with
  fermionic projected entangled-pair states},}\ }\href {\doibase
  10.1103/PhysRevB.81.165104} {\bibfield  {journal} {\bibinfo  {journal} {Phys.
  Rev. B}\ }\textbf {\bibinfo {volume} {81}},\ \bibinfo {pages} {165104}
  (\bibinfo {year} {2010})}\BibitemShut {NoStop}%
\bibitem [{\citenamefont {Phien}\ \emph {et~al.}(2015)\citenamefont {Phien},
  \citenamefont {Bengua}, \citenamefont {Tuan}, \citenamefont {Corboz},\ and\
  \citenamefont {Or{\'u}s}}]{tnref-phien15}%
  \BibitemOpen
  \bibfield  {author} {\bibinfo {author} {\bibfnamefont {Ho~N.}\ \bibnamefont
  {Phien}}, \bibinfo {author} {\bibfnamefont {Johann~A.}\ \bibnamefont
  {Bengua}}, \bibinfo {author} {\bibfnamefont {Hoang~D.}\ \bibnamefont {Tuan}},
  \bibinfo {author} {\bibfnamefont {Philippe}\ \bibnamefont {Corboz}}, \ and\
  \bibinfo {author} {\bibfnamefont {Rom\'an}\ \bibnamefont {Or{\'u}s}},\
  }\bibfield  {title} {\enquote {\bibinfo {title} {Infinite projected entangled
  pair states algorithm improved: {Fast} full update and gauge fixing},}\
  }\href {\doibase 10.1103/PhysRevB.92.035142} {\bibfield  {journal} {\bibinfo
  {journal} {Phys. Rev. B}\ }\textbf {\bibinfo {volume} {92}},\ \bibinfo
  {pages} {035142} (\bibinfo {year} {2015})}\BibitemShut {NoStop}%
\bibitem [{\citenamefont {Corboz}(2016{\natexlab{b}})}]{tnref-corboz16}%
  \BibitemOpen
  \bibfield  {author} {\bibinfo {author} {\bibfnamefont {Philippe}\
  \bibnamefont {Corboz}},\ }\bibfield  {title} {\enquote {\bibinfo {title}
  {Improved energy extrapolation with infinite projected entangled-pair states
  applied to the two-dimensional {Hubbard} model},}\ }\href {\doibase
  10.1103/PhysRevB.93.045116} {\bibfield  {journal} {\bibinfo  {journal} {Phys.
  Rev. B}\ }\textbf {\bibinfo {volume} {93}},\ \bibinfo {pages} {045116}
  (\bibinfo {year} {2016}{\natexlab{b}})}\BibitemShut {NoStop}%
\bibitem [{\citenamefont {Lohmann}\ \emph {et~al.}(2014)\citenamefont
  {Lohmann}, \citenamefont {Schmidt},\ and\ \citenamefont
  {Richter}}]{res:LSR14}%
  \BibitemOpen
  \bibfield  {author} {\bibinfo {author} {\bibfnamefont {Andre}\ \bibnamefont
  {Lohmann}}, \bibinfo {author} {\bibfnamefont {Heinz-J\"urgen}\ \bibnamefont
  {Schmidt}}, \ and\ \bibinfo {author} {\bibfnamefont {Johannes}\ \bibnamefont
  {Richter}},\ }\bibfield  {title} {\enquote {\bibinfo {title} {Tenth-order
  high-temperature expansion for the susceptibility and the specific heat of
  spin-$s$ {H}eisenberg models with arbitrary exchange patterns: Application to
  pyrochlore and kagome magnets},}\ }\href {\doibase
  10.1103/PhysRevB.89.014415} {\bibfield  {journal} {\bibinfo  {journal} {Phys.
  Rev. B}\ }\textbf {\bibinfo {volume} {89}},\ \bibinfo {pages} {014415}
  (\bibinfo {year} {2014})}\BibitemShut {NoStop}%
\bibitem [{\citenamefont {Guttmann}(1989)}]{Guttmann1989}%
  \BibitemOpen
  \bibfield  {author} {\bibinfo {author} {\bibfnamefont {Anthony~J.}\
  \bibnamefont {Guttmann}},\ }\enquote {\bibinfo {title} {Asymptotic analysis
  of power-series expansions},}\ in\ \href@noop {} {\emph {\bibinfo {booktitle}
  {Phase Transitions and Critical Phenomena}}},\ Vol.~\bibinfo {volume} {13},\
  \bibinfo {editor} {edited by\ \bibinfo {editor} {\bibfnamefont {Cyril}\
  \bibnamefont {Domb}}\ and\ \bibinfo {editor} {\bibfnamefont {Joel~Louis}\
  \bibnamefont {Lebowitz}}}\ (\bibinfo  {publisher} {Academic Press},\ \bibinfo
  {address} {London},\ \bibinfo {year} {1989})\ p.~\bibinfo {pages}
  {3}\BibitemShut {NoStop}%
\end{thebibliography}%

\end{document}